\documentclass[11pt,a4paper]{article}
\pdfoutput=1
\usepackage{jinstpub}
\usepackage[english]{babel}
\usepackage{upgreek}
\usepackage{xspace}
\usepackage{graphicx}
\usepackage[capitalize]{cleveref}
\usepackage{enumitem}
\usepackage{float}
\usepackage[T1]{fontenc}
\usepackage[utf8x]{inputenc}
\usepackage{soul}
\usepackage{xcolor}
\sethlcolor{pink}

\graphicspath{{figs/}}

\usepackage{siunitx}

\title{Design and implementation of the AMIGA embedded system for data acquisition}

\author{\includegraphics[height=30mm]{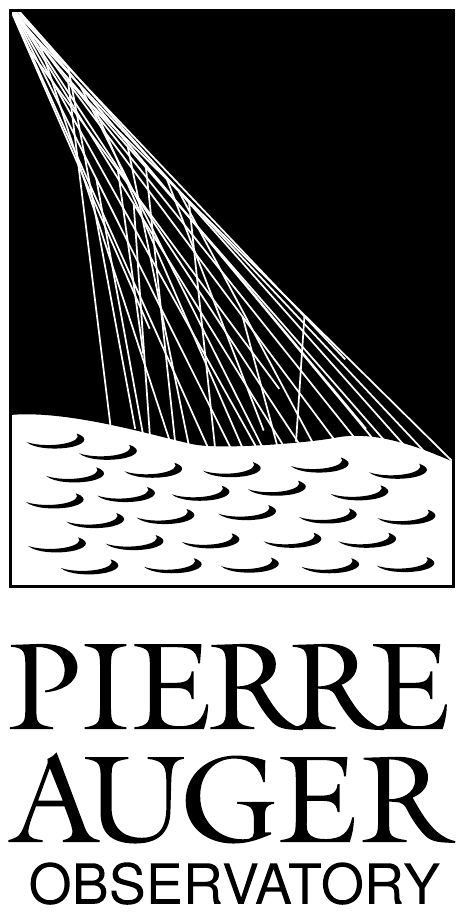}\\[3mm]The Pierre Auger Collaboration et al.}
\affiliation{Av.\ San Mart\'in Norte 306, 5613 Malarg\"ue, Mendoza, Argentina}

\emailAdd{spokespersons@auger.org}

\abstract{The Auger Muon Infill Ground Array (AMIGA) is part of the AugerPrime upgrade of the Pierre Auger Observatory. It consists of particle counters buried \SI{2.3}{\metre} underground next to the water-Cherenkov stations that form the \SI{23.5}{\kilo\metre\squared} large infilled array. The reduced distance between detectors in this denser area allows the lowering of the energy threshold for primary cosmic ray reconstruction down to about \SI{E17}{e\volt}. At the depth of \SI{2.3}{\metre} the electromagnetic component of cosmic ray showers is almost entirely absorbed so that the buried scintillators provide an independent and direct measurement of the air showers muon content. This work describes the design and implementation of the AMIGA embedded system, which provides centralized control, data acquisition and environment monitoring to its detectors. The presented system was firstly tested in the engineering array phase ended in 2017, and lately selected as the final design to be installed in all new detectors of the production phase. The system was proven to be robust and reliable and has worked in a stable manner since its first deployment.}


\arxivnumber{2101.11747}
\dedicated{\flushleft\textnormal{\textsc{Published in JINST as \href{http://doi.org/10.1088/1748-0221/16/07/T07008}{DOI:10.1088/1748-0221/16/07/T07008}
\\
Report number: FERMILAB-PUB-21-032-AD-AE-E-TD
}}}

\begin{document}
\maketitle
\flushbottom


\section{Introduction}

The Pierre Auger Observatory \cite{auger:design} was originally designed to study ultra-high-energy cosmic rays (UHECR) with primary particle energy above \SI{3e18}{\electronvolt}. It is located in the Southern Hemisphere near the Andes mountains in the south-west part of the province of Mendoza, Argentina. It uses a hybrid detection technique employing both a surface detector array (SD) and a fluorescence detector (FD).

The SD of Auger is composed of an extensive array of about 1600 water Cherenkov detectors (WCD), separated by a \SI{1500}{\metre} spacing and covering an area of \SI{3000}{\kilo\metre\squared}. As above \SI{5e19}{\electronvolt}, the flux is of around 1 particle per \SI{}{\kilo\metre\squared} per century, the large area of the observatory allows to observe more than 30 high energy cosmic rays per year \cite{auger:design}. A detailed description of the SD of the Auger Observatory can be found in \cite{auger:surface}.

Each of the 1600 surface detector stations includes a \SI{3.6}{\metre} diameter water tank containing a sealed liner with a reflective inner surface. The liner contains \SI{12000} litres of purified water. Cherenkov light produced by the passage of charged particles through the water is collected by three nine-inch photo-multiplier tubes (PMTs) that are symmetrically distributed at a distance of \SI{1.20}{\metre} from the center of the tank and look downwards through windows of clear polyethylene into the water. The surface detector station is self-contained. A photovoltaic system provides an average of \SI{10}{\watt} for the PMTs and the electronics system consisting of a processor, GPS receiver, radio transceiver and power controller.

The surface detector electronics records the PMT signals, makes local triggering decisions, sends time-stamps to the central data acquisition system for the global trigger building, and stores event data for retrieval when a global trigger condition is fulfilled. Due to the low available bandwith of \SI{150}{Bytes\per\second} per station, they must operate semi-autonomously, performing calibrations and taking action in response to alarm conditions at the station level. The electronic system was designed 15 years ago using the technology available at that time. 

\begin{figure}[ht]
  \centering
  \includegraphics[width=1.0\textwidth]{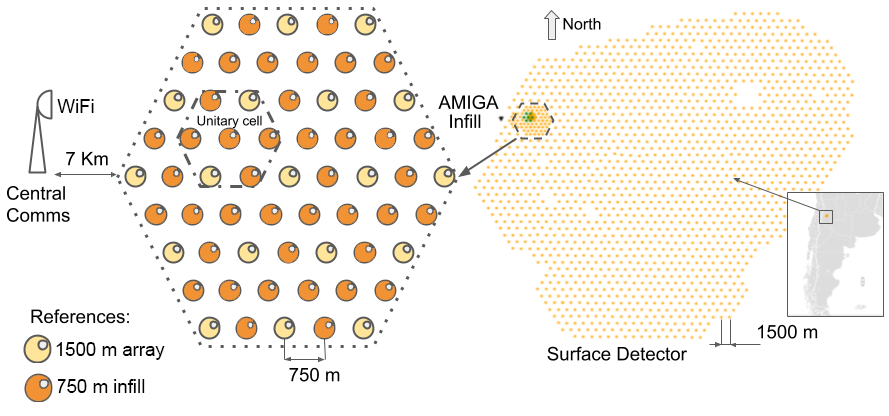}
  \caption{Map of the Pierre Auger Surface detector (right), located in Malargüe, in the southern region of the Mendoza province as shown in the lower right panel, with the AMIGA infill area marked with dotted lines and zoomed in (left).  In the blow-up the positions of the engineering array are named ``Unitary cell'' and enclosed by a dotted line. The central communication tower is located about \SI{7}{\kilo\metre} to the west (distance not to scale). The denser \SI{433}{\metre} infill is not shown in this figure for simplicity, but can be observed in figure \ref{FIG:Monit_map}}
  \label{infillAMIGA}
\end{figure}

AMIGA (Auger Muon Infill for the Ground Array) \cite{amiga:design} is an enhancement of the Pierre Auger Observatory designed to satisfy two different objectives. The first objective is to lower the minimum detectable cosmic ray energy threshold to \SI{1e17}{\electronvolt} by installing 61 SD stations with \SI{750}{\metre} spacing in an infill array covering an area of \SI{23.5}{\kilo\metre\squared}, which was completed in 2012. A more recent, and even denser array with \SI{433}{\metre} spacing over \SI{1.9}{\kilo\metre\squared} was also included and completed in 2019. Figure \ref{infillAMIGA} shows a representation of AMIGA. The second objective is to enhance the capability of Auger to study ray composition by a direct measurement of the muonic shower component. The muon detection is achieved by \SI{30}{\metre\squared} scintillator detectors buried \SI{2.3}{\metre} underground alongside all surface stations. The \SI{540}{g\per cm\squared} of overburden determined by the local soil density completely shields the electromagnetic component of the showers and only muons of energy greater than \SI{1e9}{\electronvolt} are capable of hitting the detectors.

The main function for the AMIGA Data Acquisition System is to manage the data transfer from the muon counters in a modular, flexible, easily configurable and scalable fashion compatible with the Pierre Auger Observatory data trigger. These characteristics led to specifications on how to design the firmware, software, hardware, network structure and protocol, and finally the monitoring system. For each of these important aspects of the data acquisition system, we dedicate a section on this paper, where we describe in more detail the motivation and the chosen solutions.

The AMIGA buried stations work in sync with the SD, receiving the local triggers by their accompanying WCD. The energy and geometry of the showers are reconstructed by the SD while extra information on the particle composition is provided by the muon-related observables such as the timing and size of the signal provided by buried scintillators.

In this paper we describe the features of the embedded system design for the AMIGA buried scintillator detectors, dubbed Underground Muon Detector (UMD). This description includes the electronics, as well as the synchronization with the associated SD station. As an illustration of the functionality of the whole acquisition chain, we conclude showing an event measured by the seven stations that compose the engineering array. 

The main contribution of this work to distributed embedded systems in general is the following: We present in this paper a flexible scalable massive system of multiple distributed muon detectors that can be organized in a remote fashion, using existing tools for a very specific problem: the detection of the muon component of cosmic showers. In this paper we describe how we organize the detection system using the preexisting trigger conditions of the surface detectors in the Pierre Auger Observatory and how we can apply reliable existing technology that can be adapted and easily applied as a solution to this problem. The electronics system of the front-end is described in a companion paper \cite{AMIGA:INTEGRATOR} where the details of the front-end electronics of the UMDs, its design and performance are described. In summary, we must mention that the UMDs work as muon counters, using two types of detection: 1 bit multichannel digitalization, one for each of the 64 channels, and a {\SI{14}{bit}} Analog to Digital Converter with two independent channels that measure the charge of the signal from the 64 channels for a muon count that exceeds 64 muons.


\section{Underground Muon Detector overview}

As the main objective of the UMD is to count the number of muons in a cosmic shower produced by a cosmic ray, the original design of AMIGA was based on particle detectors that would be able to count individual muons in a radiation detector. This detector was implemented as a segmented scintillator, divided into three modules of 64 channels each, with a total area of detection determined via simulations \cite{AMIGA:sim}. These simulations were also used to obtain the optimum detector segmentation (\SI{192}{} scintillator bars). The total area needed of \SI{30}{\metre\squared} was therefore covered by three modules of \SI{10}{\metre\squared} each.

All SD stations in the infilled array have been installed and are operating since 2008 (61 stations) \cite{amiga:first}. The SD stations in the infilled \SI{750}{\metre} array have been installed since 2008 (42 additional stations) \cite{amiga:cell} and the complete grid is operative since 2012. More recently, the SD 433m array was also completed. The companion UMD had an engineering array (EA) phase of seven stations to validate and optimize the detector design. This prototype phase ended in 2017 with two major changes: the replacement of the optical device from PMTs to SiPMs and improved electronics, which is presented in this work. Immediately after the EA phase, the production phase started and the UMD array completion is foreseen by the end of 2022 \cite{amiga:cell}.

AMIGA uses a telecommunications system that consists of a point-to-multipoint 802.11n WiFi radio link, with redundant coordinators located at the Coihueco fluorescence detector building (about \SIrange{6}{10}{\kilo\metre} link distance). That link provides TCP/IP remote access to the LAN of each station \cite{amiga:comms}. The system is also designed to be energy efficient (as power from the photovoltaic system is limited) and has the capacity to transmit a high amount of data from those remote areas.
All AMIGA stations are expected to work at least 10 years in remote (and possibly hard to reach) areas. Under these conditions, the systems need to have a long service life and to allow remote access for diagnostics and early fault detection for high quality of data.

The electronics were designed as embedded systems to ease eventual hardware upgrades. They run Linux as open source operating system to be free from any particular commercial software and be able to implement different hardware choices used on the detectors.

\subsection{Electronics Overview}

The electronics for the UMD was designed to provide a reliable data transfer from the buried scintillator modules to the main data storage server in Malarg\"ue. In order to do this, each UMD is connected to an SD to use the Pierre Auger Observatory trigger. In order to provide a safe data transfer and not to overload the existing SD telecommunications and power system, each station was equipped with a new telecommunications system \cite{amiga:comms}, a new solar power system \cite{amiga:power}, a trigger relay system (the \emph{distributor}, see section \ref{AMIGA:DISTRIB}), three buried modules that comprise the UMD itself, and a synchronization system that connects the UMD data acquisition system with its corresponding SD data acquisition system. The motivation for the telecommunications system is stated in the mentioned reference, as a summary, an 802.11n WiFi system running a TCP/IP network is used. The motivation for the solar panel system design is stated in the mentioned reference. In summary, a \SI{24}{\volt} power system was implemented to provide the power needed via two deep discharge \SI{12}{\volt} batteries connected in series that satisfy the power requirements of the UMDs and their telecommunications system, a total of \SI{\sim 46}{\watt} in the engineering array for at least one day without solar energy (For production the power of the buried electronics of the UMDs was decreased to \SI{\sim 3.6}{\watt} as measured in the field, reducing the total power budget to \SI{\sim 19}{\watt} per station in the main array, allowing the photovoltaic system to have an extended autonomy of at least three days without solar energy). The trigger distribution and synchronization mentioned was motivated by the need to provide muon counting data stored in the UMDs each time a trigger request from the main data processing center in Malarg\"ue is received. Therefore, the UMDs must be synchronized with the SDs and the trigger lines must be distributed to each of the three buried muon counters in each station.

\begin{figure}[htb!]
  \centering
  \includegraphics[width=0.9\textwidth]{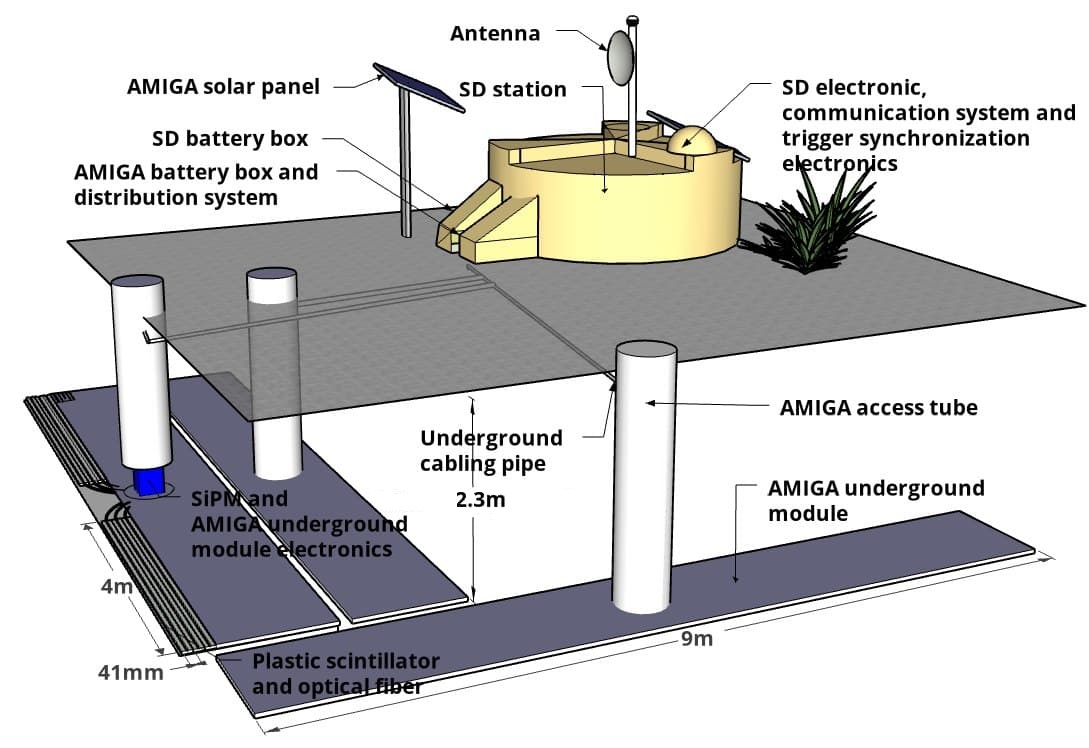}
  \caption{General overview of an infill station. The figure shows a counter and surface detector. Three modules are shown buried \SI{2.3}{\meter} deep, with a detection area of \SI{10}{\metre\squared}. Module electronics are accessed for field installation and service via a plastic tube. In addition, the solar panel and the additional battery box are shown.}
  \label{amiga_station}
\end{figure}

A drawing that illustrates the design of the SD and UMD combined detectors at a position in the infill is shown in figure \ref{amiga_station} with its main components. Three \SI{10}{\metre\squared} buried modules are used at each surface detector station for the final design \cite{AMIGA:sim}.

\begin{figure}[htb!]
  \centering
  \includegraphics[width=0.8\textwidth]{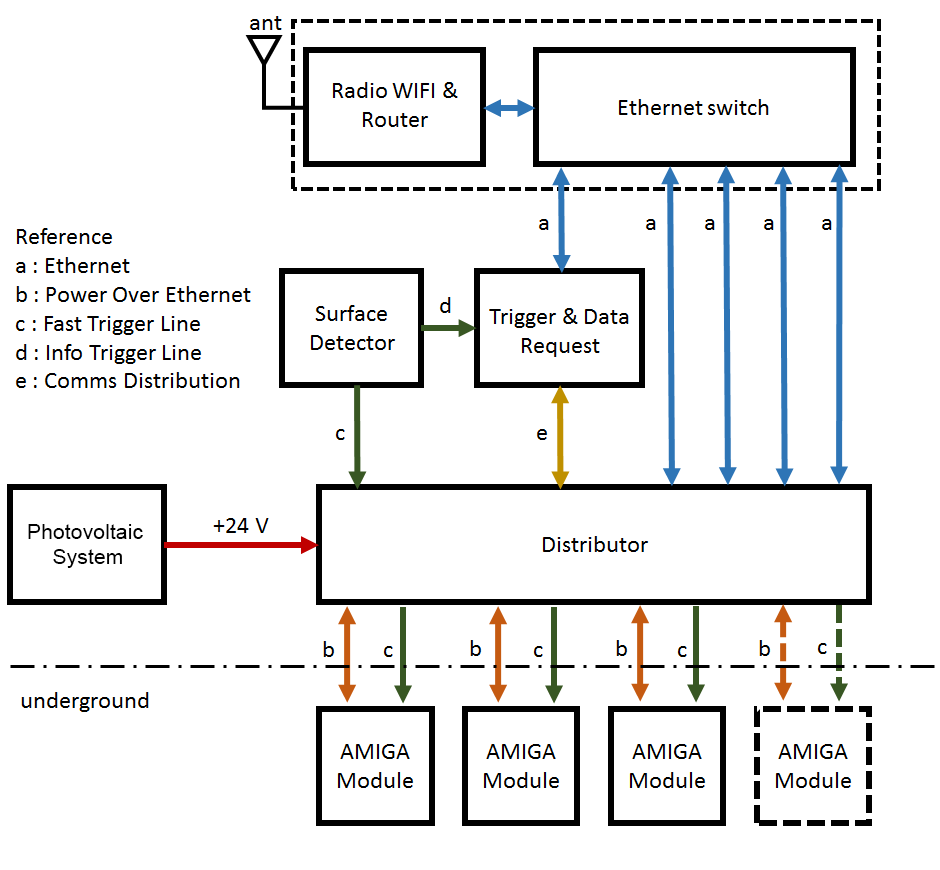}
  \caption{Block diagram of an AMIGA station electronics. The Ethernet switch conforms the center of the star of the Ethernet LAN. The distributor replicates the fast trigger signal from the SD station and injects power from the photovoltaic system to the Ethernet cables that connect the buried modules.}
  \label{amiga_diagram}
\end{figure}

The block diagram of the electronics is shown in figure \ref{amiga_diagram}, which describes the components of one combined station and its interconnection as detailed in the following subsections.

\subsection{UMD scintillator module}

In this section, we present a summary of the underground scintillator module and its electronics functionality. A thorough description of the scintillator module and the buried electronics has already been published in \cite{AMIGA:prototype}, \cite{AMIGA:SiPM} and \cite{AMIGA:INTEGRATOR}. In summary, the motivation for the design of this electronics was to provide a ring buffer that serves as a temporary storage big enough to save muon data of a full cosmic shower for the time the Auger Observatory trigger takes to send a request for that data. In this section we introduce the scintillator module, the hardware of the underground electronics, the structure of the stored data and the requirements for its implementation in the digital back-end of the underground electronics. We also review the hardware structure of the buried electronics interconnected with the scintillator bars using optical fibers.

Every module comprises 64 scintillation bars, each of \SI{41}{\milli\metre} width, \SI{10}{\milli\metre} thickness and \SI{4}{\metre} length for a detection area of \SI{10}{\metre\squared}. The light produced in each scintillator bar is collected and propagated along a wavelength-shifting (WLS) optical fiber of \SI{1.2}{\milli\metre} diameter glued into a lengthwise groove of the bar. All fibers connect to an optical coupler (also called ``cookie'') located in front of the light sensor. The \SI{64}{} scintillators and optical fibers are enclosed within a PVC (Polyvinyl Chloride) casing and they form together with the electronics the detector module. A detailed description of the AMIGA detector is found in \cite{AMIGA:prototype}.

Two groups of \SI{32}{} scintillator bars are mounted in each module at opposite sides of a central dome that contains the light sensor and electronics. For the final design we selected the Hamamatsu S13361 Silicon Photomultiplier (SiPM) \cite{AMIGA:SiPM}. An integrated electronics acquires the analog signals from this SiPM, processes them and provides control, monitoring and communication functionality to the module. The system records event data synchronized with the associated SD station (at about \SI{100}{events\per\second}) and stores them for roughly the same time as the SD station. Each recorded event is composed of \SI{2048}{} samples of \SI{64}{bits} measured at \SI{320}{Msps} and \SI{1024}{} samples of two \SI{14}{bit} ADCs (high and low gain) measured at \SI{160}{Msps} \cite{AMIGA:INTEGRATOR}. The former and the latter are dubbed the {\it binary} and the {\it integrator} acquisition modes respectively. Then, an event with a temporal length of \SI{6.4}{\micro\second} requires: \SI{64}{bits} x \SI{2048} + 2 x \SI{16}{bits} x 1024 = \SI{163}{Kbits}. Notice that the \SI{14}{bits} traces are sent in two Bytes, hence the \SI{16}{bits} value in the second term. Added to this trace are a header of \SI{336}{bits}, a trace of 2 x \SI{2048}{bits} that correspond to the OR gates in the two CITROCs of the front-end \cite{AMIGA:INTEGRATOR} and \SI{16}{bits} x \SI{2048} that are the identifiers of each the bins in the \SI{64}{bits} trace. All this adds up to a total number for the trace of: \SI{163}{Kbits} + \SI{336}{bits} + 2 x \SI{2048}{bits} + \SI{16}{bits} x \SI{2048} = \SI{201}{Kbits}. In order to store the data of a candidate event for about \SI{20}{\second} at a rate of \SI{100}{events\per\second}, the minimum memory needed is \SI{402}{Mbits}.

The electronics for each module consists of the front-end board and the Acquisition and Control board. The front-end processes the 64 light signals by two Application Specific Integrated Circuit (ASIC) \cite{AMIGA:SiPM} and two \SI{160}{Msps} \SI{14}{bit} ADCs (Analog Devices ADS4246) digitizing the sum of all SiPM channels (one with a low-gain amplifier and one with a high-gain amplifier). Each ASIC channel provides a pre-amplifier with programmable gain and a fast shaper with \SI{15}{ns} peak time. Finally, a discriminator digitizes the signal into one bit. The discriminator threshold has a coarse setting via a \SI{10}{bit} DAC (common for the 32 channels) and a per-channel fine setting using individual \SI{4}{bit} DACs. The Acquisition and Control Board (also called in this paper the \emph{back-end}) records the digital information of the front-end board and stores the data in a ring buffer when a trigger condition is found. Details are given in section \ref{sec3}.

\subsection{Synchronization} 

In order to comply with the trigger structure of the Pierre Auger Observatory, we need to establish how the synchronization of the UMD will be carried out whenever a cosmic shower is detected by the SDs. The detailed description of the trigger structure of the observatory is already thoroughly described in \cite{auger:trigger} and \cite{auger:design}. In this section we outline the main restrictions our Data Acquisition System has to abide by in order to comply with these trigger requirements. The specifications described in this section serve as a reference for the trigger interconnection between the surface and buried electronics, the memory usage in the ring buffer and the data handling that has to be performed in the SD in order to successfully transfer muon data from the UMD to the main server in Malarg\"ue that is effectively synchronized with the detection of a cosmic shower.

The Auger surface detector data acquisition and storage system is comprised of a central storage server (or ``CDAS'' located in the city of Malarg\"ue) and several client stations; the muon counter replicates this structure, merging the muon counter data with the surface detector data in a post-processing step offline.

The trigger system of SD is hierarchical: The first level of trigger (T1) is generated independently by each station based on local analysis of the signals produced by its PMT. At calibrated SD stations the threshold is set for an average T1 trigger rate of about \SI{100}{events\per\second}. These events are stored in a circular buffer with a capacity of \SI{3000}{events}. Events are identified by a 64-bit time-stamp (GTS) obtained from the GPS receiver with a latency of about \SI{200}{\micro\second} after T1 as measured in the laboratory. The station then applies predefined quality cuts to generate a second level trigger T2 with an average rate of \SI{20}{Hz}. A list of T2 time-stamps is sent to CDAS once every second. CDAS searches through the list of T2 triggers for a coincidence in vicinity and time. When a coincidence is found, an array trigger (T3) is sent back to participating stations and its neighbors. Upon receiving a T3 trigger request, the station replies to the CDAS with the event data, if it is present in the circular buffer \cite{auger:trigger}, \cite{auger:design}.

UMD buried modules were designed to rely on an external trigger in order to acquire event data during a possible particle shower. To provide synchronization with the underground particle counter modules, additional electronics were installed in the surface station. These complementary electronics were designed to trigger and label the particle counters data when the surface detector encounters a T1 trigger condition, and to forward T3 data requests from CDAS \cite{design_report} to the buried electronics afterwards.

\subsubsection{Trigger output from Auger Surface Detectors}

The time that precedes a T1 event strongly depends on the delay between the issuing of the trigger signal and the arrival of the identifier. To address this issue a buffer that stores the T1 data must take this delay into account. Also a faster additional time-stamp was implemented, capable of arriving at the buried modules within a few microseconds after the SD station detects the trigger condition. This time-stamp, called Local Time-Stamp (LTS) consists of a \SI{24}{bit} number generated by the surface station. Taking into account the average T1 rate, each LTS value is guaranteed to be unique for at least \SI{26.8}{\second}, enough time to ensure that requested events will not have time-stamp collision (and therefore data collision) with newer events \cite{amiga:digital}. The time it takes between the detection of an event condition (T1) and the sending of the signal is about \SI{760}{\nano\second}. However, using this time-stamping mechanism requires a way of converting between the time-stamp used by the surface detector and the LTS. This conversion is provided by electronics named Trigger and Data Request (TDR).

\begin{figure}[ht]
  \centering
  \includegraphics[width=0.8\textwidth]{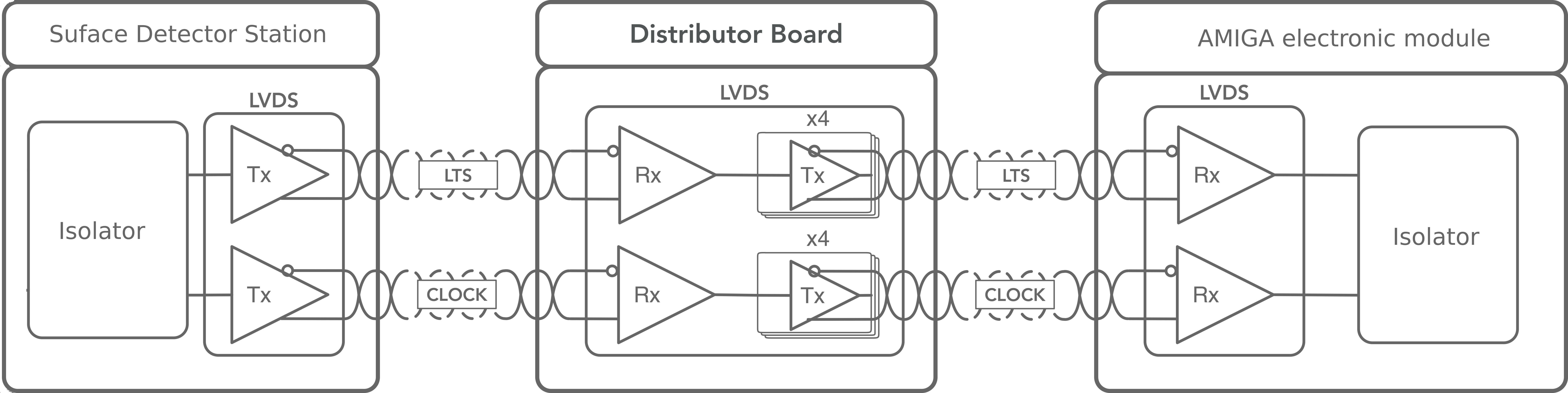}
	\caption{T1 circuit (LTS) between the surface detector electronics and the modules. The simplified circuit for a line is shown in order to have a better understanding.
	The implemented circuit distributes the signals in the distributor, after the LVDS receiver. }
  \label{fig:circuitT1}
\end{figure}

Both trigger signals (to the buried electronics and the information sent to the TDR) are sent from an auxiliar board retrofitted in every SD station with muon counter detectors installed (``Trigger-TX''). The minimum required data transmission rate is \SI{10}{Mbps} (\SI{100}{\nano\second} per bin). This board has two low-voltage differential signalling (LVDS) drivers. Each LVDS line uses a point-to-point configuration, using drivers and receivers from Fairchild Semiconductors (FIN1001 and FIN1002 respectively) supporting data rates higher than \SI{600}{Mbps}.

The typical power consumption of the transmitter is \SI{23}{\milli\watt} and of the receiver \SI{13}{\milli\watt}, thus each data line requires \SI{36}{\milli\watt}. This system uses two of the available four pairs of lines to send signals (data and clock), which gives a power consumption per link of \SI{72}{\milli\watt}.

In order to avoid ground loops, the fast trigger signal is isolated before it reaches the LVDS driver using an Digital Isolator from Texas Instruments (ADuM1400), as shown in figure \ref{fig:circuitT1}. The transmission line is terminated by a \SI{100}{\ohm} resistor placed very close to the receiver for best impedance match. An Unshielded Twisted Pair (UTP) Cat 5e cable with RJ45 connectors crimped at both ends (following TIA/EIA norm 586-A) is used to transmit the signal.

The delays (calculated from timing data in each component's datasheet) are shown in table \ref{tab:delay}.

\begin{table}[ht]
\centering
\begin{tabular}{|c|c|c|}
\hline
Component & Delay & Unit \\ \hline
Isolator     & \SI{43}{}      & \SI{}{\nano\second}     \\ \hline
Driver LVDS & \SI{ 5}{}       & \SI{}{\nano\second }    \\ \hline
Cable UTP   & \SI{ 6}{}       & \SI{}{\nano\second\per\metre}   \\ \hline
LVDS receiver   & \SI{ 5}{}       & \SI{}{\nano\second}   \\ \hline
\end{tabular}
\caption{Summary of delays for components involved in signal transmission.}
\label{tab:delay}
\end{table}

Figure \ref{fig:circuitT1} shows that the signal crosses two isolators, two LVDS transmitter / receiver pairs and two sections of UTP cables. With a \SI{6}{\metre} cable between SD electronics and distributor and a \SI{30}{\metre} cable to the UMD electronics the latency for signal transmission is \SI{322}{\nano\second}.

To calculate the total delay, the time needed for the acquisition system to recognize the trigger condition must be added to the signal latency calculated below. This time is about \SI{500}{\nano\second} therefore, the trigger of the AMIGA electronics will arrive about \SI{1600}{\nano\second} after the trigger signal has been generated in the Auger station electronics.

\subsubsection{Trigger and Data Request}

In the beginning of the engineering array phase the functionality of the TDR was realized by a Single Board Computer (SBC) with a buffered SPI receiver on the basis of an Altera MAX-II complex programmable logic device (CPLD). Since the production phase, an AMIGA Acquisition board is used for this purpose. With the installation of the new SD electronics called ``Upgraded Unified Board'' (UUB) \cite{UUB} as part of the Auger upgrade, the TDR functions are available in this board and no auxiliary electronics are needed.

The TDR receives the GTS+LTS time-stamp from the Auger surface electronics via a \SI{10}{Mbps} SPI channel and sends it to the SBC, which stores these time-stamps in a circular buffer with a depth of \SI{2048}{} values. The TDR wiretaps the T3 requests to the station by a RS-232 interface at the moment. The UUB has all the hardware ports, software (T3 trigger to the buried electronics and distributor commanding/monitoring), and FPGA code needed to replace the TDR. Therefore, on stations with a UUB installed, the TDR and SPI line are removed, and the distributor control cable is connected to the UUB directly.

When a T3 is found, the TDR looks for the received GTS in the GTS+LTS table. If the GTS is found, its corresponding LTS is broadcasted to all the modules via UDP; otherwise, a ``LTS not found'' message is sent.
The TDR also provides access to the SD serial console via SSH or TELNET protocols. Note that this trigger scheme has no dependency on the buried module electronics, and can be reused by any project designed to trigger with Auger surface detector stations.

\subsection{Communications}
The telecommunications system connects the AMIGA stations to the central storage server, and it comprises of an Ethernet LAN network (connecting the UMDs and the SD electronics) and a WLAN telecommunications system for the data transmission to CDAS. The WLAN is implemented as a point-to-multipoint WiFi radio link attached to the original Auger inter-FD microwave links and it is provided by a Mikrotik RB493 router, which includes an Ethernet switch and a 802.11n WiFi radio. These characteristics are mentioned in \cite{amiga:comms}, where we can also find that the network structure for the UMDs was implemented over the \SI{2.4}{GHz} WiFi band. The motivation for this decision was to have a modular system that is higly reliable to work outdoors and that can handle the throughput required for the network of detectors over the \SI{2.4}{GHz} Wi-Fi band, for which the Pierre Auger Observatory has a licence for exclusive use in the local region granted by the \emph{Comisi\'on Nacional de Comunicaciones} of Argentina. The details of the design, the protocol as well as the tests performed to measure its reliability are described in the mentioned reference. In this section we expand the description of the network to the higher layers and how it is organized in subnetworks within each detector.

\subsubsection{Network}
The network addressing scheme for communications uses one /24 network (Class C network, 253 available IP addresses) for each station LAN, with all radios connected on the same WLAN network. (\texttt{172.16.0.0/16}). A station with an internal LAN network \texttt{10.a.b.0/24} will have the IP \texttt{172.16.a.b} in the WLAN network.
In the internal detector LANs, the IP \texttt{10.a.b.1} is reserved for the gateway (WiFi radio), \texttt{10.a.b.2} for the TDR, and \texttt{10.a.b.(10 + x)} for the module x of that detector (when x $<$ 90). 



\subsection{Power and Trigger distribution}
\label{AMIGA:DISTRIB}

This section explains the distribution of the Pierre Auger Observatory triggers and their interconnection at a physical level with the UMDs. It is worth mentioning that the trigger and power distribution between a SD and the three buried UMDs are intimately related since they share the same lines, therefore we also include a brief description of the power and monitoring system used. As mentioned before, the main piece of hardware that is in charge of handling the trigger and power lines to the buried UMDs is called the \emph{distributor}, and in this section we introduce its functionality.

The power for each station is supplied by independent photovoltaic systems since all installation sites are far away from any electric power grid \cite{amiga:power}. The distributor receives power from the solar panels and distributes it to the four modules via Passive Power-over-Ethernet (PPoE) lines. It uses the same cable to distribute the fast trigger signal sent to the buried modules.

The distributor converts the LVDS signals from the Auger surface electronics to TTL-CMOS logic using LVDS receivers. The TTL fast trigger is isolated in order to avoid a possible ground loop and is split into four lines. Each line is independently converted to LVDS signals, available at the distributor output ports. Each PPoE output uses the pin-out of 802.3af mode B\footnote{\SI{+24}{VDC} on pins 4 and 5 and GND on pins 7 and 8} and is connected to the power supply through a relay that allows switching the power on and off. The relays are commanded by the TDR using RS-232 messages transmitted via UTP cable, allowing up to four distributors in a daisy-chain (required when more than four counters are present in a station).


\section{Acquisition and Control System}
\label{sec3}
In this section we present the functional structure of the Data Acquisition and Control (DAQ) system of the UMDs. We describe the implementation at the hardware level on a Cyclone IV FPGA, the timing issues and memory requirements for the tasks performed by the DAQ system, including the firmware of each UMD and the soft microprocessor platform used to implement every function required by the UMD. We also describe the power system distribution and consumption of the electronics, the interface required and finally a short description of the printed circuit board layout and how it was designed. As we will see in this section the constraints of speed and memory are significant, therefore we emphasize the need for a soft microprocessor platform in order to have easy access to upgrades or future hardware changes that may be required due to spares availability. In this way we have a flexible and scalable embedded system that can be used for any kind of particle or radiation detector that works synchronously. The Acquisition and Control System is composed by two distinct parts: A soft microprocessor and custom acquisition code (both detailed below). A soft microprocessor was chosen to simplify the electronics, reducing the component count and improving the electronics future-proofing.

\subsection{Hardware}

In this section we describe the back-end digital hardware used for data acquisition on each UMD. We introduce the basic structure of the main system, comprised of an FPGA and a memory that functions as a ring buffer storing data according to a trigger provided by the SD. The main CPU controls the interface and data transfer to the surface, as well as monitoring several key hardware parameters that indicate its good performance. As the main motivation for the design, we should mention the need for a compact flexible embedded system that can easily be reprogrammed remotely, as we will see in the following sections. As mentioned, the CPU and firmware were implemented using an FPGA connected to two physically independent LPDDR1 memories. This board has an expansion port with \SI{150}{} pins using a \SI{1.8}{V} single ended voltage standard for fast speed digital signals. Additionally, one RJ-45 connector receives the fast trigger (T1) and the clock line with LVDS receivers, another one uses a LVDS transmitter/receiver pair for serial communication. These LVDS drivers can be fed by power from the connected cable, or by power from the Acquisition board, selected by a jumper. All input/output signals from LVDS transmitters/receivers are electrically isolated. In the following subsections we describe the hardware used with its specifications in terms of timing, resources and memory usage, the printed circuit board design and the interface with the SD and the telecommunications system.

\begin{figure}[ht]
  \centering
  \includegraphics[width=0.8\textwidth]{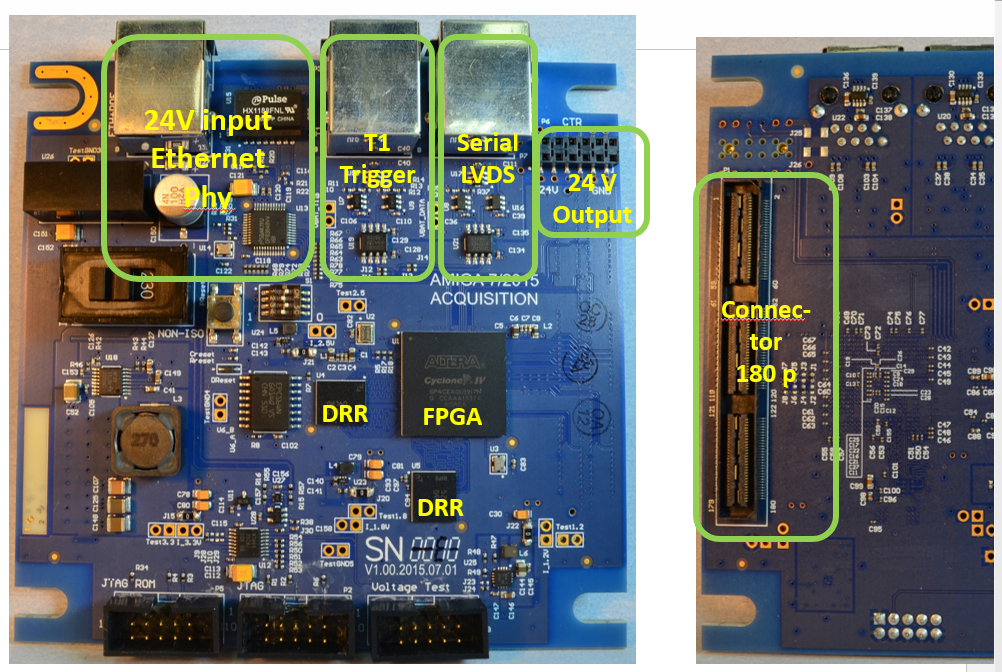}
  \caption{Acquisition board, showing the main connectors and components of the top side (left), and the front-end connector located at the bottom (right).}
  \label{FIG:FullBoard}
\end{figure}

Figure \ref{FIG:FullBoard} shows a picture of the top and bottom side of the assembled board. It depicts the layout of the board and indicates the locations of the main connectors and components.

\subsubsection{FPGA and memory}

The board is built around an Altera Cyclone IV FPGA (model EP4CE40U19I7) which provides \SI{39600}{} logic cells and a memory of \SI{1.16}{Mbits}. Table \ref{TAB:LUTCount} summarizes the usage of these resources for the custom logic (Firmware) and the LEON3 softcore CPU. While only \SI{18}{\%} of the logic cells are not used, there is nearly half of the memory cells (\SI{48}{\%}) left for future improvements.

The chosen industrial package with \SI{329}{} I/O pins (of which \SI{257}{} are used) works in a wide temperature range ($-40~^\circ$C to $+100~^\circ$C), which is adequate for the operating conditions of the electronics (buried underground).
The FPGA series is optimized for low power consumption with a low core voltage of \SI{1.2}{\volt}. The firmware incorporates the Altera Remote Update system \cite{ARU} with a remote recovery option, which ensures additional safety in case of an upgrade failure.

\begin{table}[ht]
\centering
\begin{tabular}{|c|c|c|}
\hline
\textbf{Component} & \textbf{Logic Cells} & \textbf{Memory bits} \\ \hline
Firmware    & 12156 (\SI{30}{\%})  & 480211 (\SI{41}{\%}) \\ \hline
LEON3 CPU   & 20371 (\SI{51}{\%})  & 133872 (\SI{11}{\%}) \\ \hline
Total       & 32527 (\SI{82}{\%}) & 614083 (\SI{52}{\%}) \\ \hline
Available   & 39600 (\SI{100}{\%}) & 1161216 (\SI{100}{\%}) \\ \hline
\end{tabular}

\caption{Resource usage of the FPGA image, split into LEON3 and Firmware. These numbers were obtained from logs generated by the the Quartus II compiler.}
\label{TAB:LUTCount}
\end{table}

The board has two physically separated memories, one for the soft-core CPU and one for the acquisition firmware. This design avoids logic elements for the synchronized access to the two RAMs, thus speeding-up the acquisition firmware and saving resources. Both memories are Micron LPDDR (low power DDR, model MT46H32M16LFBF-5), with \SI{1}{GB} of capacity and a working frequency of up to \SI{200}{MHz} (Both memories are used at \SI{100}{MHz}).

\begin{figure}[ht]
  \centering
  \includegraphics[width=0.6\textwidth]{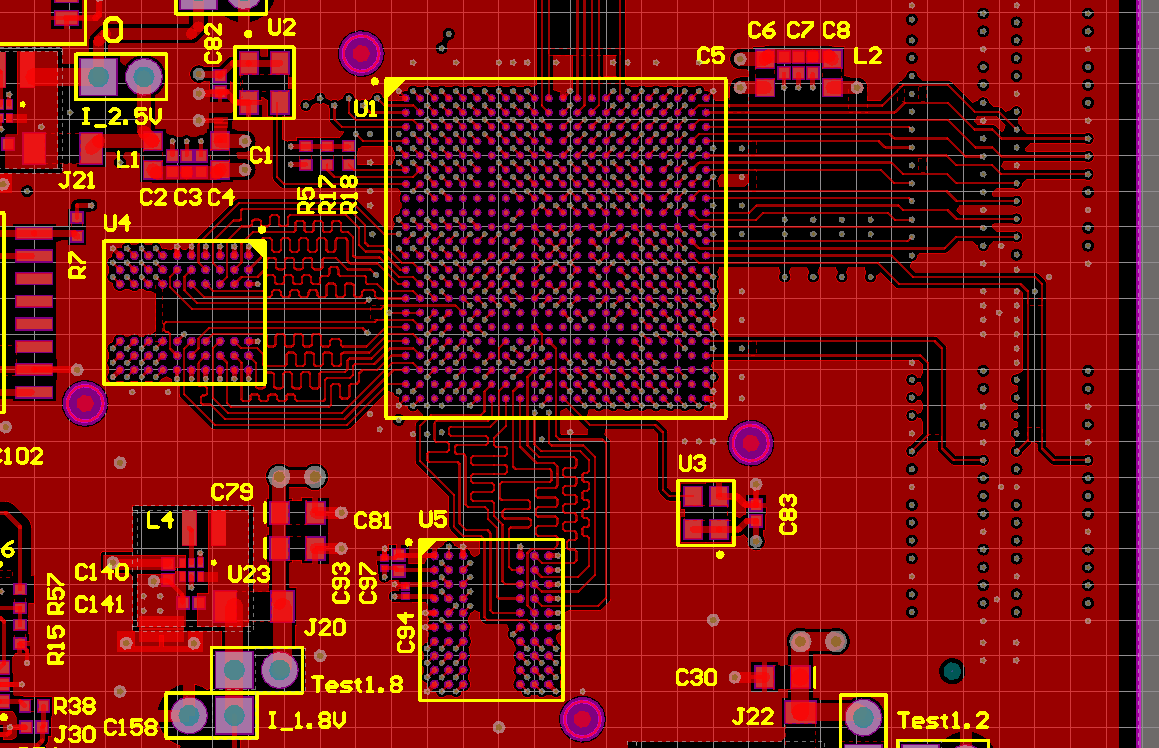}
	\caption{Close-up of the PCB Layout focusing on the FPGA (``U1'') and LPDDR memories (``U4'' and ``U5'').}
  \label{FIG:PCBLAYOUT}
\end{figure}

\subsubsection{Power sources}

\begin{figure}[ht]
  \centering
  \includegraphics[width=0.4\textwidth]{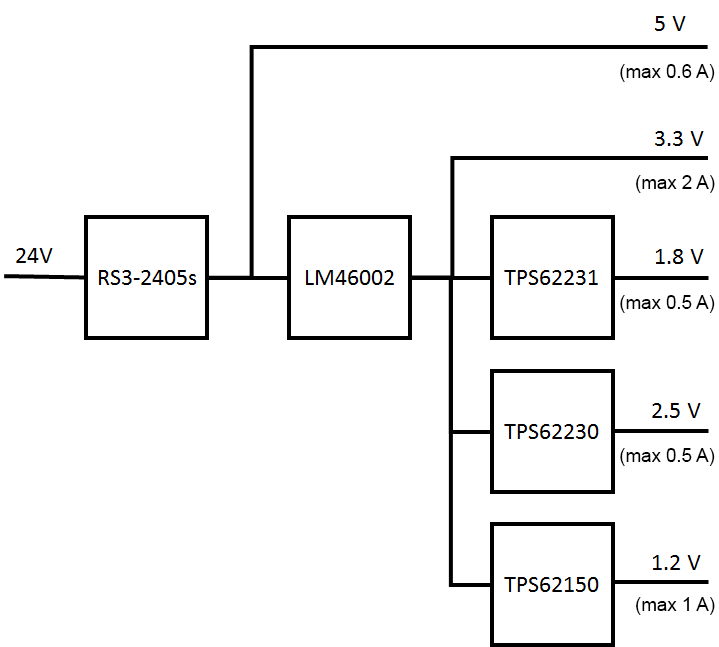}
  \caption{Power Supply diagram. Indicated are the maximum power and current supported for each component. }
  \label{FIG:PWRSUP}
\end{figure}

Based on the estimates for the required ratings of the different devices, we designed a cascaded high-efficient power system as shown in figure \ref{FIG:PWRSUP}.

\begin{table}[ht]
\centering
\begin{tabular}{|r|r|r|}
\hline
\multicolumn{1}{|l|}{\textbf{Voltage [V]}} & \multicolumn{1}{l|}{\textbf{Current [mA]}} & \multicolumn{1}{l|}{\textbf{Power [W]}} \\ \hline
24      & 75    & 1.800   \\ \hline
3.3     & 210   & 0.693   \\ \hline
1.8     & 44    & 0.079   \\ \hline
2.5     & 46    & 0.115   \\ \hline
1.2     & 236   & 0.283   \\ \hline
\end{tabular}
\caption{Power consumption for each of the outputs shown in figure \ref{FIG:PWRSUP}.}
\label{TAB:PWRConsumo}
\end{table}

The board uses a \SI{3}{W} isolated DC/DC-converter RS3-2405S to avoid ground loops. The device converts the nominal \SI{24}{V} input into a \SI{5}{V} output (max \SI{0.6}{A}). This output is converted by the LM46002 DC/DC-regulator to a stable \SI{3.3}{V} supply (max \SI{2}{A}). With this \SI{3.3}{V} as input, further step-down DC/DC regulators TPS62150, TPS62230 and TPS62231 generate \SI{1.2}{V} (max \SI{1}{A}), \SI{2.5}{V} (max 0.5A) and \SI{1.8}{V} (max \SI{0.5}{A}) with an efficiency of typically \SI{95}{\%}, respectively. The \SI{3.3}{V}  and \SI{1.8}{V}  power supplies are connected to the FPGA I/O blocks, providing power and reference voltage. Table \ref{TAB:PWRConsumo} show the nominal values of current and power consumption as measured in the laboratory for each of the outputs in figure \ref{FIG:PWRSUP}.

\subsubsection{Interface and Communications}

The board provides two expansion connectors: one for data/logic signals and one for power supplies. The data expansion connector is a single high-speed connector, Samtec QSH-090-01-LDAK-TR, capable of transmitting signals with a maximum frequency of \SI{9}{GHz}. It also has a central ground bar that ensures low resistance for the return currents through the connector. The connector has 180 lines, with a pitch between pins of \SI{0.5}{mm}.

The power expansion connector provides the \SI{24}{V} input power (directly from the PPoE connector) and the \SI{3.3}{V}  power supply. Additionally, two control lines provide the functionality to switch on or off the connected expansion board.

The Ethernet PHY driver is a DP83848IV from Texas Instruments, which is supported by the Leon 3 built-in Ethernet MAC and the Linux operating system. The MAC address of each board is divided in two parts: The most significant 11 nibbles (44 bits) are a firmware compile-time constant, and the lowest nibble (4 bits) can be set using four switches in the board.

For the monitoring of the system voltages an ADS7828 ADC from Texas Instruments was chosen. It has 8 multiplexed input channels, uses an internal reference voltage, and supports I2C communications for transferring monitoring data to the soft-core.

\subsubsection{Printed circuit board}

Based on the maximum working frequency within the board (\SI{\sim 200}{MHz}) it was decided that the appropriate material for this design is FR4, which can accommodate working frequencies of up to \SI{500}{MHz}. A 12-layer design was made, following the recommendations of the FPGA and RAM manufacturers. The FPGA recommendations specify tracks with a minimum width of \SI{5}{mils}  and a characteristic impedance of \SI{50}{\ohm}. The impedance of the lines depends on the materials (dielectrics and conductors) and their geometry, that is, track width, thickness of the dielectric, among others. A standard thickness (\SI{1.6}{mm}) FR4 PCB with 12 layers does not allow \SI{50}{\ohm} lines, therefore a thicker (\SI{2.4}{mm} ) PCB was used.

Following the application notes of Altera, one LPDDR memory was connected to a whole FPGA I/O bank. The other memory was connected using part of a bank for the address pins and part of another for the data bus, as indicated by the FPGA manual \cite{CycloneIVDDR}. Once this was defined, the routing of the tracks from the FPGA to the memories was done.

The tracks between RAM and FPGA must be striplines of controlled impedance and length to conserve signal integrity and the timing constraints imposed by the RAM. The track width and distance between the signal and reference layers needed for a specific impedance was determined by specialized programs taking into account the properties of the PCB material. They must also have equalised lengths to ensure that the propagation times of the signals are within the values recommended by the RAM manufacturer.

\subsection{Soft microprocessor LEON3}

The back-end design went through several iterations before arriving to the final design presented in this paper. Through this debugging process we concluded that the flexibility of the hardware was essential, as well as the remote control and configuration. After two prototypes were built, we chose to use a LEON3 soft microprocessor running at \SI{50}{\mega\hertz} that has an open source architecture. It allows for the system to be fully programmable remotely from the Observatory campus in Malarg\"ue, which eased its debugging on the field and allows for upgrades or changes in the firmware to accommodate future needs such as changes in the trigger system, calibration, monitoring and disabling malfunctioning parts of the hardware if necessary. LEON3 is a 32-bit CPU microprocessor core, based on the Scalable Processor Architecture Version 8 (SPARC-V8) instruction set architecture with an AMBA AHB system bus \cite{LEON3}. It was originally designed by the European Space Research and Technology Centre (ESTEC), part of the European Space Agency (ESA), and after that by Cobham Gaisler Research. It is described in synthesizable VHDL.

This Soft microprocessor has numerous advantages, particularly for a such long-running project: The base code (CPU without Floating Point support and most peripherals) is available under a free-software license (GPLv2), it is FPGA-independent (upstream code supports Altera, Xilinx and Microsemi FPGAs), comes with a complete library of peripherals, and allows a wide array of configuration options (both in adding and removing peripherals and selecting or removing CPU features to prioritize performance or resource usage). The peripherals used are: SPI Memory controller (used for early boot and remote Firmware upgrade), I2C (used for environmental sensors), RS-232 serial ports (used for High Voltage power supply and acquisition control/data transfer), GPIO ports (used for the front-end power supply control, reboot/power-off and watchdog), and 10/100 Ethernet MAC (used for communications).

\subsection{UMD firmware}

\begin{figure}[ht]
\centering
\includegraphics[width=0.8\textwidth]{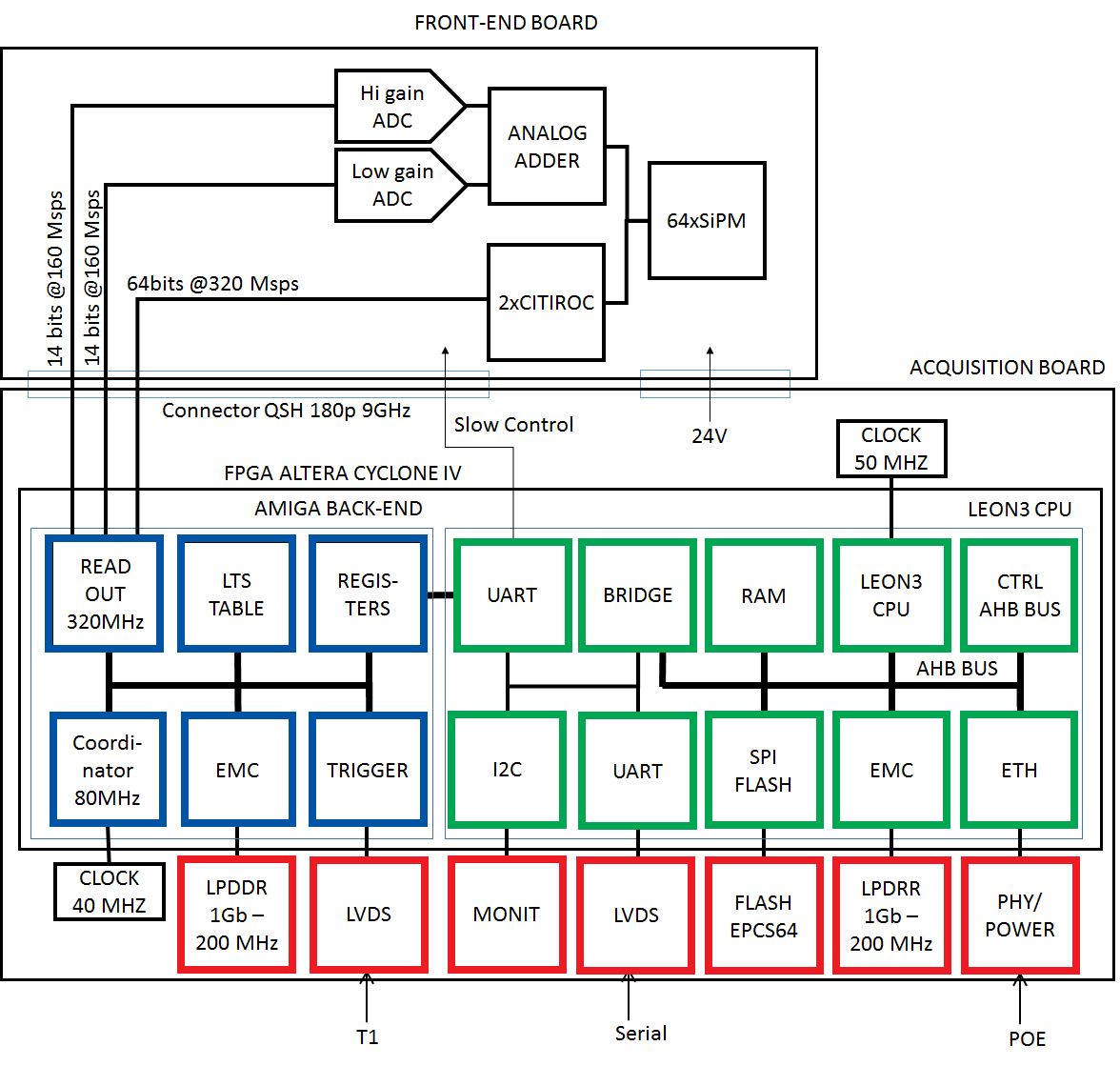}
\caption{Block diagram of the UMD electronics. We can observe in the acquisition board: in green the CPU blocks, in blue the back-end blocks and in red the peripheral blocks.}
\label{FIG:BMEBlockDiagram}
\end{figure}

In this section we introduce the details of the firmware functionality for a UMD back-end. It handles the data flow and secondary tasks required of the UMD hardware. We describe each part of the firmware in a block diagram divided into three parts: the CPU core, the back-end main task of muon data storage and the interface with the front-end and the peripherals. The current FPGA implementation is divided into six main blocks, as shown in figure \ref{FIG:BMEBlockDiagram}. The EMC block drives \SI{128}{MiB} of external LPDDR1 memory for temporary event data storage. This data has a fixed size (\SI{2048}{} anode samples for each SiPM, plus data from both ADCs and the aux digital outputs from both ASICs, equivalent to \SI{6.4}{\micro\second} per event). The trigger block receives the T1 signal, immediately notifies the incoming event condition to the data block and forwards the LTS value corresponding to the incoming event once it has been fully received. The read-out stage processes the \SI{64}{} digital inputs at \SI{320}{\mega\hertz} and packs four samples into a single \SI{256}{bit} word. To reduce the number of resources needed, in particularly interconnects, all firmware except for the readout process runs at \SI{80}{\mega\hertz}. The data acquisition block stores \SI{2048}{samples} of these \SI{256}{bit} words in two alternating circular (internal) buffers. When a trigger arrives, the actual buffer is filled up and then switched to the alternative buffer. While the alternate buffer records the data, the recorded data is saved in the external RAM. The LTS table process manages a look-up table with 2048 entries. When addressed with an LTS of an event, the memory address of the event data block is returned, or an `event not found' error is raised \cite{amiga:digital}.

The ASIC programming block receives the ASIC configuration from the embedded system and then uses it to program both ASICs. The micro-controller interface block handles communications with the embedded system. This is implemented using serial ports. Finally, the coordinator block manages the rest of the blocks.

\subsection{Remote upgrade}

In order to have a flexible distributed system we need to provide a robust firmware upgrade system. Therefore we included in the main firmware a remote upgrade that can be performed by an operator and provide a failsafe if there is any problem during the upgrade. In this way we can remotely enhance and develop the main functions of the UMD. Altera Remote Upgrade works as follows: The hardware stores two FPGA images in a persistent memory, named ``Factory'' and ``Application'' in Altera literature. On FPGA power-on, the Factory image is programmed, which starts a watchdog and immediately programs the Application image. If the watchdog isn't periodically reset (at least once every 60 seconds), the Factory image is reprogrammed in the FPGA and takes over. Both firmwares can be identified by software via a GPIO pin. The Application image contains both a LEON3 and AMIGA code, while the Factory image contains only a LEON3. The watchdog is controlled by software, and the routines to reset it start late in the boot process. If the module boots in Factory mode, the software tries to re-enter Application mode ten minutes after booting has completed.

Firmware upgrades are done by writing a specially processed FPGA image into the SPI Flash. To prevent an overwriting of deployed software images by mistake, all Flash regions except the one containing the FPGA Application image are set write-protected.


\section{Software overview}

This section describes how the software in the UMD is implemented, the tasks provided to the UMD and the operating system used. The software is responsible for sending event data and monitoring data via a network to a remote location. To simplify and speed-up the development phase, this software runs over an operating system which also provides system tools. The system was designed to minimise the use of persistent storage in the module electronics, improving resiliency, easing electronics installation and replacement, and reducing the possibility of ``bricking'' the electronics on future software updates (the only persistent data on every module electronic is LEON3, Firmware, boot-loader and Ethernet MAC address). A description of the software, the boot process and the SDK follows.

\subsection{Operating System}

Following the open source philosophy adopted for the soft microprocessor, the operating system used is also open source. The underground scintillator detector modules and the TDRs use Linux \cite{linux} as the operating system. It was chosen for user familiarity, relative ease of use, and because it has full hardware support for LEON3 (CPU and required peripherals). Therefore the motivations for implementing the software in this manner is to facilitate remote upgrades and configuration, with a complete access to the hardware functionality of the UMD electronics deployed in the field. The current version used at the time of the writing of this paper is 4.9, with patches (provided by Cobham Gaisler) for hardware support forward-ported to 4.9. Additionally, the kernel was patched to use the built-in GPIO power-off driver, which has been-setup to instruct the FPGA to program the Application image. The system runs in RAM from an initcpio (embedded in the kernel image due to bootloader limitations). Loadable module support is disabled (all modules are compiled into the kernel image) because no out-of-tree kernel modules are used, the connected hardware is known in advance, and disabling module support results in smaller kernel and initcpio binaries.

\subsubsection{Boot}

Upon board power-on/CPU Reset, a boot-loader (uBoot 1.16 with Cobham Gaisler patches) starts running. This boot-loader brings up the Ethernet port, obtains an IP address via DHCP, and loads the full operating system image (specified by the DHCP reply) from a TFTP server located in the Mikrotik RB493 router in each station. The DHCP server runs on the central data storage server, with DHCP relays running on each station LAN. The TFTP servers are located in each station LAN to preserve WLAN bandwidth. This system allows for easy image upgrades: Simply overwrite the image on the respective TFTP servers and reboot the affected modules. Modules are assigned a fixed IP and hostname based on their MAC address. This address has the upper 44 bits set by the software image (common for all stations) and the lower 4 bits configurable by hardware-switches, admitting at most 16 module electronics per station. Since each station LAN is physically isolated and has its own network address, no MAC address conflicts arise as long as MAC addresses are not repeated within the same station LAN.

\subsubsection{System Applications}
All basic system applications (init, shell, etc) are provided by Busybox (currently v1.33.1) \cite{busybox}. The most notable applications are: Telnet daemon, watchdog user-space application, NTP client, DHCP client and SPI flash. The DHCP client has been configured to obtain the client hostname published by the DHCP server (required by AMIGA software to identify the module), the station LSID (used by the monitoring software to identify the station), and the NTP server used to synchronize the internal time.

\subsubsection{Software Development Kit (SDK)}
The tool-chain used for software development (compiler, libc, linker and related utilities) was custom-built. It consists of a GCC 8.1 C compiler targeting by default the Leon3 architecture with software floating point support, using binutils 2.30 and uCLibc-ng 1.30 C library. The tool-chain was built for Ubuntu 18.04 running on Intel x86\_64 compatible CPUs.

\subsection{Software}
In this section we describe in more detail the software structure and how it performs the tasks needed for the UMD to store data and transfer it to the CDAS. We organize this section into three main subsections that describe the most important routines and their functionality: Module configuration, data transfer, and calibration. The software was written to control the module and to manage event and monitoring data acquisition. It consists of five pairs of client-server applications. With the single exception of the Calibration software (\texttt{MdCalib}), all clients run automatically after booting in Application mode on each buried module electronics and all servers run automatically on boot on a server in CDAS, with the clients keeping a persistent connection to the server (unless specified). Additionally (and with the same exception), all programs use the host-name to identify each individual module, allowing the system to work behind a Network address translation (NAT). A brief description of each pair of programs, with the names of each program client/server side, follows.

 \begin{figure}[ht]
  \centering
  \includegraphics[width=0.9\textwidth]{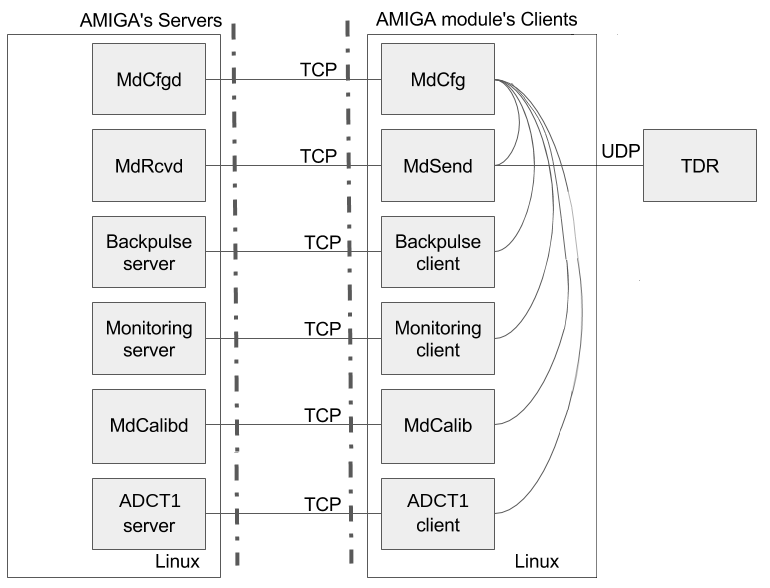}
  \caption{Software communications diagram.
           All communications with the central server go through TCP sockets, maximizing data integrity and minimizing data loss.}
  \label{soft}
 \end{figure}

\subsubsection{Module configuration - \texttt{MdCfg}/\texttt{MdCfgd}}
This software configures the front-end (CITIROCs, ADCs, and HV), the firmware parameters and provides configuration/status information (ex: acquisition mode, masked channels, etc) to other programs using a Unix socket. On start-up, the client connects to the server and requests the initial module configuration (stored in the central server in INI files). Once the configuration data has been sent and applied, this daemon disconnects from the server and keeps running in background, handling configuration and status requests from all other programs (these requests will be specified in each program's section).

\subsubsection{Data transfer - \texttt{MdSend}/\texttt{MdRcvd}}
This pair of programs handle the event data transfer and storage. When idle, the client periodically consults with \texttt{MdCfg} the current acquisition mode and the UDP broadcast port for data requests (only if in acquisition modes sending of data is requested). Upon a data request, the software instructs the firmware to search for the specified event, receiving either the event data or an error code, and sends the result to the server. The server keeps in RAM a dictionary of events indexed by the 16-bit internal CDAS Event ID. For each Event ID, the data for all participating counters is stored in a dictionary indexed by the counter ID. The server has a TTL counter for each event, which is decreased every 30 seconds and increased every time data for a new module arrives.
Once this counter reaches zero, the event is written to disk and discarded from RAM (if more data for the same event arrives, it is treated as a new event and merged as part of the standard offline post-processing of AMIGA data). The event data is stored in JSON format, using a single file which is periodically restored and compressed (in gzip format) once per day. This format allows easy parsing using libraries available in most programming languages, and has a very good compression ratio when using gzip \cite{JSON}.

\subsubsection{Calibration - \texttt{MdCalib}}
These programs run the pre-defined Calibration routines on each module as described in \cite{AMIGA:SiPM}. On the module, \texttt{MdCalib} starts on boot and listens to calibration requests. Once a calibration request is received, the program acquires the file-lock for background data, sets the acquisition mode to `Calibration' (all data requests are sent to the server indicating that the sent data is invalid due to the module being calibrated) and runs the requested calibration routine, setting the needed configuration parameters via \texttt{MdCfg} and sending the resulting background data to the requester. Once the calibration finishes, the module is returned to its previous state and the file-lock is released. On a remote machine, a dedicated program communicates with the calibration software running on one or more modules (identified by IP address) and receives the calibration data. Currently the request for calibration date must be issued manually by an operator, but full automation (via scheduled jobs) can be used.

\subsection{Monitoring system}
\label{monitoring}

To guarantee the validity of the acquired data, the status of the detector, as well as the status of its measured data, have to be monitored. In this section we include a description of the monitoring system as implemented in the UMD and the main server. The motivation for the monitoring system is described below, in summary it is designed to perform two very important tasks: Background signal monitoring, where the record of the signal counts by the UMD is constantly taken, and environmental monitoring, where the hardware status (including power sources voltages, currents and temperature) is handled. The monitoring acquisition system was designed as a series of processes (one for each kind of monitoring) using client/server architectures, where each of the clients runs in a MC electronic and establishes a TCP connection to its corresponding server. The resources needed by the monitoring processes are extremely low, particularly from the client side: they use \SI{1}{MB} of RAM that is activated every 5 minutes to send data to the server at the CDAS. The data volume sent is \SI{1}{KByte} every \SI{30}{minutes}. A description of these monitoring processes follows.

\subsubsection{Background signal monitoring - \texttt{backpulse\_client}/\texttt{backpulse\_server}}
These programs read and store the background pulse counters from the firmware. Background pulses are all the current pulses produced by the SiPM (not necessarily belonging to a possible event). The rate of these pulses for a given SiPM is expected to be relatively constant throughout the operating lifetime of the detector; variations on this rate over time may indicate aging or damage of any of the optical parts. The client periodically orders the firmware to snapshot the current counter values, which are then sent to the server and saved in space-separated files (one file per day and module), formatted with the module time-stamp and the 64 counter values. Since the background data client shares resources with the calibration daemon, this program tries to acquire a lock-file in a predetermined location before reading the counter data, and immediately closes this file after reading.

\subsubsection{Environmental monitoring - \texttt{mo\-ni\-to\-ring\-\_cli\-ent/mo\-ni\-to\-ring\-\_ser\-ver/da\-ta\-\_han\-dler}}
This trio of programs handle the monitoring data readout and storage. The client periodically reads the data directly from the monitoring ADCs and data from the HV power supply via \texttt{MdCfg}, and sends it to the monitoring server upon request. Monitoring data is taken every \SI{5}{minutes}. The server spawns a new process for each connection request from a client which periodically requests its monitoring data. Every \SI{30}{minutes} up to \SI{432}{Bytes} of monitoring data plus a header of \SI{35}{Bytes} is sent per station. This data is then immediately sent to the \texttt{data\_handler} process via a unix socket. The \texttt{data\_handler} process stores all the monitoring data it receives from its unix socket to a MySQL database and, as a backup, to a .csv file (rotated every \SI{12}{hours}) that can be imported manually into the database.

\subsection{Online-Monitoring}

At the server side, the monitoring data is received and stored in a database. In this section we describe the monitoring data handling from all the UMDs deployed in the field and how it is presented to the user in order to visualize it and make decisions accordingly.

\begin{figure}[ht]
  \centering
  \includegraphics[width=0.5\textwidth]{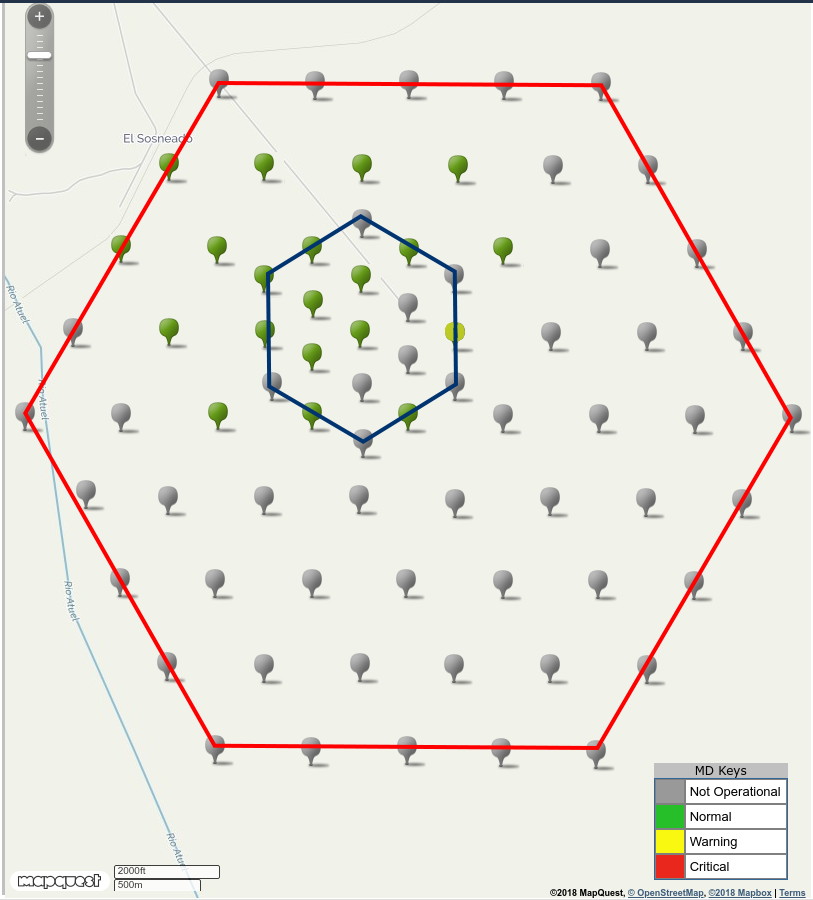}
  \caption{Screen capture of the monitoring system (integrated within the Auger monitoring system) displaying UMD stations. Green ticks indicate stations with buried modules in normal operating status. The red hexagon circumscribes the \SI{750}{\metre} infill and the blue hexagon circumscribes the \SI{433}{\metre} infill. Displayed are the stations deployed at the time of publication of this paper. The station in yellow that shows a warning is displayed in detail in figure \ref{FIG:PeterMazur}.}
  \label{FIG:Monit_map}
\end{figure}

The monitoring system allows access to the most relevant information of the different components of the MC via a web-interface. It also enables the visualisation and analysis of the environmental and trigger monitoring parameters over time, which can be very useful to find anomalies that could indicate the presence of failures in some component of the MC.

\begin{figure}[ht]
  \centering
  \includegraphics[width=0.9\textwidth]{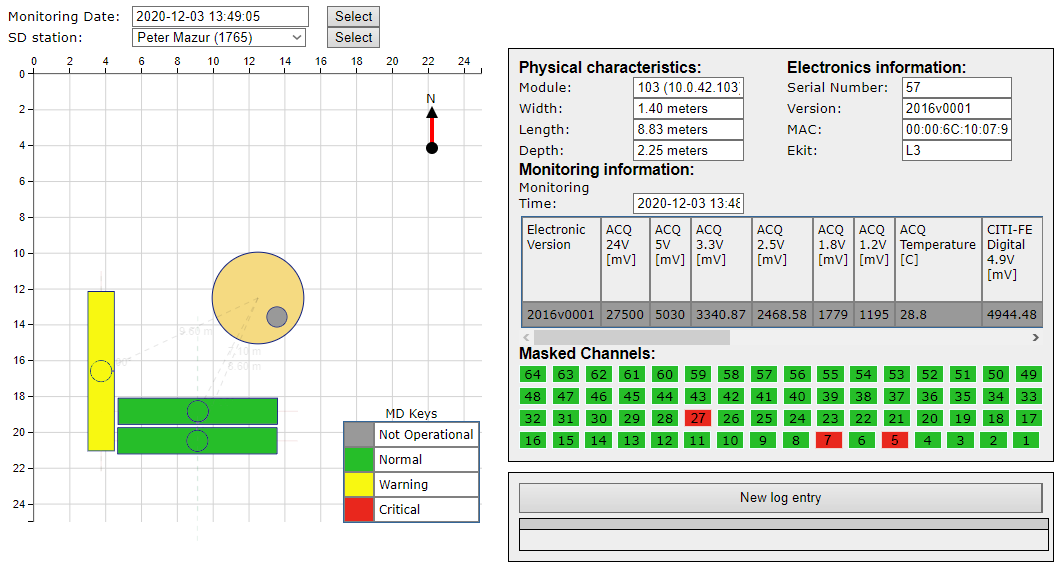}
  \caption{Screen capture of the monitoring system, displaying installation and monitoring data for an UMD. As can be observed, module 103 shows a warning as three of its channels (5, 7 and 27) are not operational.}
  \label{FIG:PeterMazur}
\end{figure}

The main page of the monitoring system shows a Map option with the position of each SD station. Figure \ref{FIG:Monit_map} shows an example, where the coloured marker depends on the alarm state of all the scintillator modules for each SD station.
In the figure, green marks represent a station working in normal state, and gray marks represent stations that have yet to be installed.

Selecting a SD station shows the deploy view associated to this station and the monitoring data from the database for a particular date. An example of the monitoring system displaying a typical AMIGA station after deployment is shown in figure \ref{FIG:PeterMazur}. Selecting a module shows the monitoring information and alarm state for each variable of the chosen module. This view also gives the option to mask alarms of any monitoring parameters and visualise the full monitoring history of the corresponding module.


\section{AMIGA sample event}

\begin{figure}[ht]
  \centering
  \includegraphics[width=0.9\textwidth]{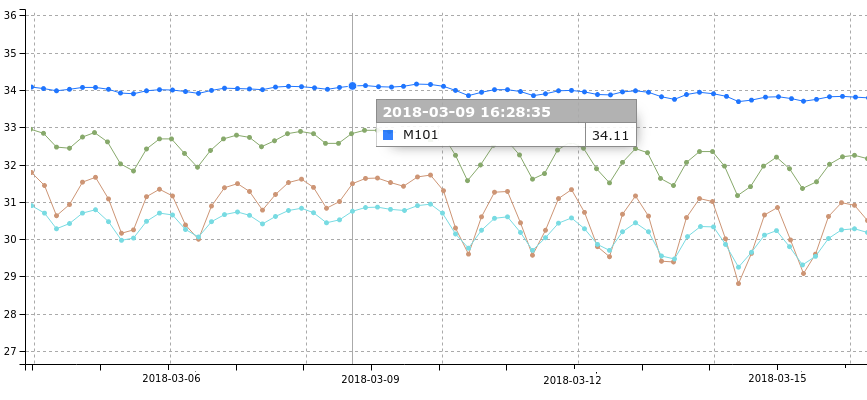}
	\caption{Time plot of the SiPM Temperature Sensor (in \SI{}{\celsius}) for four AMIGA modules of station with ID 1764. The 12-day time period includes the moment of the presented event (at the solid vertical line).}
  \label{FIG:1764_South_Temps}
\end{figure}

As an example of the acquisition process, this section presents data collected by the AMIGA UMDs for an UHECR of \SI{4e18}{eV} impinging with a \SI{20}{}$^\circ$ zenith angle\footnote{Values taken from the official Pierre Auger Observatory event reconstruction}, that hit the border of the muon counters.
This event was recorded on May 9, 2018 at 14:44:56 UTC. The temperature measured in the SiPMs for the four modules belonging to station 1764 (which participated in the event) is shown in figure \ref{FIG:1764_South_Temps}, taken form the monitoring system. 
Event data for each AMIGA station 
is shown in this section to illustrate the reconstruction procedure applied to the {\it binary} acquisition mode only\footnote{As the time of writing this paper, the reconstruction with the {\it integrator} mode is still under development}.

\subsection{Status of the AMIGA Array at the time of the event recording}

\begin{figure}[ht]
  \centering
  \includegraphics[width=0.8\textwidth]{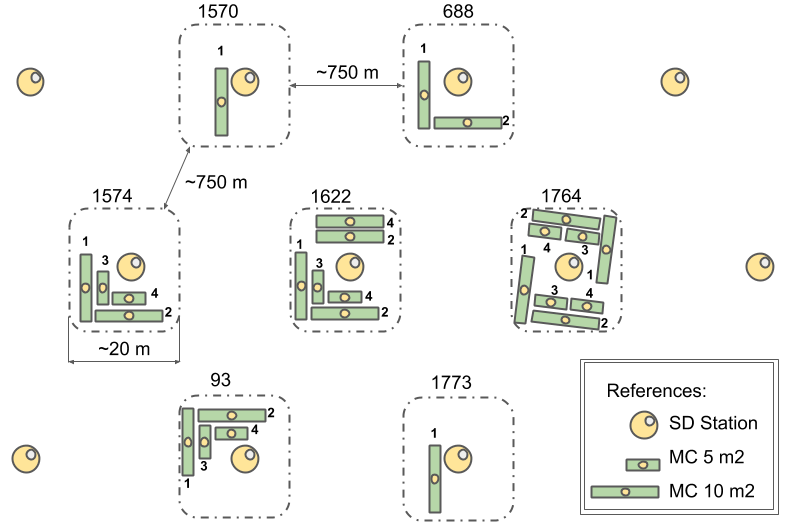}
  \caption{Status of the AMIGA Engineering Array at March 2018. Each station is identified by a number (``Station ID''). For each station the size and relative orientation of the AMIGA muon counter is sketched. The size of the modules represent the the detection area (\SI{10}{\metre\squared} or \SI{5}{\metre\squared}). Drawing is not to scale.}
  \label{AMIGAUnitaryCell}
 \end{figure}

As seven stations of the \SI{750}{\metre} array were part of the prototyping phase (the so called {\it Unitary Cell}), the first operative hexagon for physics analysis has stations with higher segmentation (256 bars instead of 192) for the projected \SI{30}{\metre\squared} detection area as shown in figure \ref{AMIGAUnitaryCell}.
Two of these stations have four buried modules (two \SI{10}{\metre\squared} modules and two \SI{5}{\metre\squared} modules).
One of these stations has eight modules (four \SI{10}{\metre\squared} modules and four \SI{5}{\metre\squared} modules) installed symmetrically with respect to the Auger Surface Detector; also known as a ``Twin'' (for validation purposes and to assess the resolution of the reconstruction procedure as described in \cite{MuonsWithAMIGA}).
One station has six modules (four \SI{10}{\metre\squared} modules and two \SI{5}{\metre\squared} modules), with the southern side of the station installed like the others, and the northern side with two \SI{10}{\metre\squared} modules.
The rest of the stations have one or two \SI{10}{\metre\squared} modules.

\subsection{UMD module binary data}

AMIGA event data for each buried scintillator module is stored as lists of bin number/anode sample, only for samples where at least one channel has detected light (all other samples are not stored), and a list of bin number/sampled value for each ADC.
The anode bin number indicates a time-stamp in \SI{3.125}{ns} increments.
These traces have a length of 2048 bins (\SI{6.4}{\micro\second}).
The length of the pre-trigger / post-trigger window is a configurable parameter of the electronics. In this example, the pre-trigger and post-trigger windows are about \SI{4.4}{\micro\second}/\SI{2}{\micro\second} long, thus the trigger arrived at bin \SI{1400}{}.

\begin{figure}[ht]
  \centering
  \includegraphics[width=0.80\textwidth]{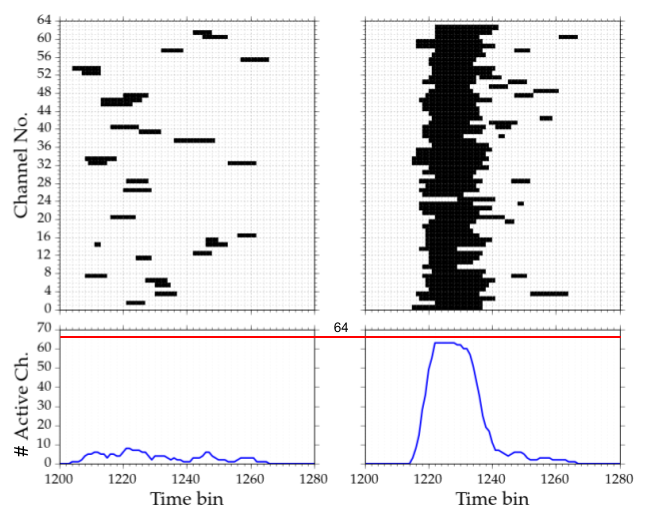}
	\caption{Example of acquisition for two  modules in two different stations. 
 (left) Plots of acquired event data for the presented event on a non-saturated AMIGA buried muon counter module (1622 south M1) with 29 pulses. (right) Plots of acquired event data for the presented event on a saturated AMIGA buried moun counter module (1570 south M1) with 64 pulses.
The top graphic shows a black rectangle on each channel that had signal over threshold in a given time bin.
The bottom graphic shows the sum of all channels with signal over threshold for a given time bin. The red line indicates the level when all 64 channels are active. A station is considered to be saturated based on the amount of active channels for a time bin.
}
  \label{FIG:trace_1622_101}
\end{figure}

Muon counts are obtained from these raw traces (shown as illustration in top panels of figure \ref{FIG:trace_1622_101}) in an offline analysis, searching for patterns in the signals of each channel using the technique described in \cite{AMIGA:SiPM}. 
A counting strategy must be applied to avoid undercounting from pile-up effects \cite{AMIGA:sim}.

As an example of the reconstruction chain applied to extract a number of muons from the raw data, figure \ref{FIG:trace_1622_101} shows the traces collected by two AMIGA modules at different distances from the shower core for the same event. It is apparent how the particle density and therefore
the triggered raw traces decreases as the distance increases. In particular, the figure shows an extreme case for a very close-to-core detector at \SI{145}{\metre} distance. In this case, the UMD is saturated\footnote{In the {\it binary} mode} as all 64 channels in three of its modules are simultaneously triggered.
To identify the number of active channels, {\it i.e.} the number of channels with at least four positives samples in the raw traces in a time window of 18 bins, is the first step in the reconstruction procedure \cite{AMIGA:SiPM}. 
For the station 1570, only one of the larger modules (M1) of \SI{10}{\metre\squared} is not saturated and measures 471 particles. On the other hand, for the \SI{10}{\metre\squared} M1 (south) buried module of the detector 1622 (at about \SI{600}{\metre} of the shower core) the active channels are 29. Applying the pile-up correction, this number of active channels translate into 33.7 reconstructed muons.
This correction is applied to the number of active channels, and makes this number differ from the reconstructed muons.
The method relies on a probabilistic model to estimate the number of particles, reporting the mean number of muons for a given
number of input signals (or active channels).

As a final remark, it is important to stress that the detected particles arrived in a time window of about \SI{200}{\nano\second} and that there are no identified particles before or after this window. 

The reconstructed number of muons for all AMIGA modules triggered in this sample event is presented in table \ref{TAB:EVENT_NUMBERS}. These counts were obtained using the Auger Offline software \cite{PAO:offline}, which applies the already described counting strategy and the pile-up correction described in \cite{AMIGA:sim}.

\begin{table}[ht]
\centering
\begin{tabular}{|r|r|r|r|r|r|r|}\hline
\multicolumn{1}{|c|}{Station} & \multicolumn{1}{|c|}{\begin{tabular}[c]{@{}c@{}}Distance \\ to axis (m)\end{tabular}} & \multicolumn{1}{|c|}{$\mu$ density $[\frac{\mu}{\SI{}{\metre^2}}]$} & \multicolumn{1}{|c|}{\begin{tabular}[c]{@{}c@{}}M1\\ (\SI{10}{\metre\squared})\end{tabular}} & \multicolumn{1}{|c|}{\begin{tabular}[c]{@{}c@{}}M2\\ (\SI{10}{\metre\squared})\end{tabular}} & \multicolumn{1}{|c|}{\begin{tabular}[c]{@{}c@{}}M3\\ (\SI{5}{\metre\squared})\end{tabular}} & \multicolumn{1}{|c|}{\begin{tabular}[c]{@{}c@{}}M4\\ (\SI{5}{\metre\squared})\end{tabular}} \\\hline
1570 S    & 145      & 44.86         & 471.1        & -            & -               & -           \\\hline
1622 S    & 603      & 2.69          & 33.7         & 29.4         & 10.1            & 11.3        \\\hline
1622 N    & 603      & 2.84          & 34.7         & 25.0         & -               & -           \\\hline
688 S     & 633      & 3.17          & 41.8         & 24.8         & -               & -           \\\hline
1574 S    & 669      & 2.10          & 19.7         & 25.0         & 9.2             & 12.2        \\\hline
1764 S    & 1114     & 0.41          & 4.0          & 5.0          & 3.0             & 1.0         \\\hline
1764 N    & 1114     & 0.44          & 6.1          & 3.0          & 1.0             & 4.0         \\\hline
93 N      & 1170     & 0.22          & 1.0          & 2.0          & 1.0             & 3.0         \\\hline
1773 S    & 1344     & 0.00          & 0.0          & -            & -               & -           \\\hline
\end{tabular}
\caption{Muon counts calculated for the presented event, separated by station, and analysing twin detectors separately. }
\label{TAB:EVENT_NUMBERS}
\end{table}






\begin{figure}[ht]
  \centering
  \includegraphics[width=0.7\textwidth]{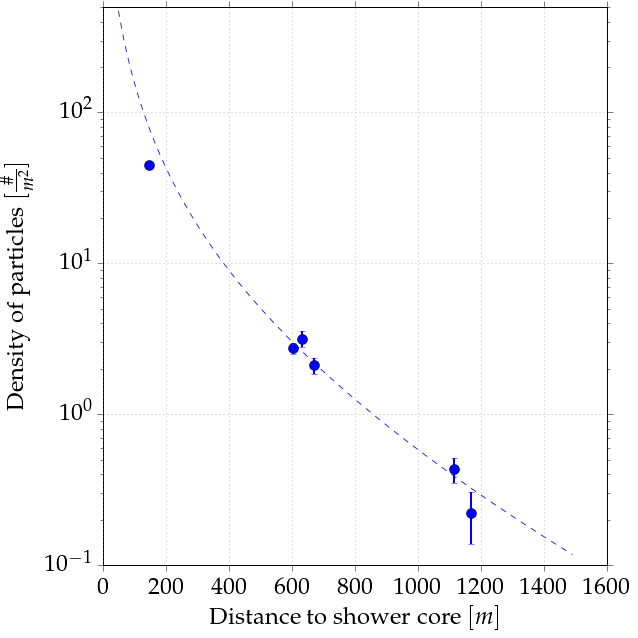}
  \caption{MLDF (Muon Lateral Distribution Function) of the studied event, including a fit (dashed blue line) of the calculated data. This event has data points for distances to the core smaller and greater than \SI{450}{m}, which results in a better quality fit.}
  \label{FIG:MLDF}
\end{figure}

After the muon densities as a function of the distance are reconstructed, the observations are fit to obtain the
estimator $\rho_\mu(450)$, the muon density at \SI{450}{m} from the shower core. This observable is sensitive to the chemical composition of the primary cosmic ray entering the Earth atmosphere and is the one used for higher-level physical analysis. 
Combining this parameter with others from the shower reconstruction (as the energy, X$_{\textrm{max}}$, etc), gives valuable information regarding the primary particle mass.
Figure \ref{FIG:MLDF} shows the fit 
of the muon lateral distribution function made for the event studied 
following the procedure in \cite{Ravignani2015}.



\section{Conclusions}

The AMIGA front-end electronics was designed as an embedded system around the Cylone FPGA family. A soft-core LEON3 processor runs Linux as its operating system for high flexibility, short development time and other benefits. A front-end compatible with this board was designed and built for AMIGA buried muon counters using SiPMs as the sensing device and two CITIROC ASICs for signal conditioning. The system is capable of 64 SiPM channels sampled at frequencies of \SI{320}{Msps} and two \SI{14}{bit} ADCs (high and low gain) measured at \SI{160}{Msps}.

The first eight prototypes of this system were installed in October 2016.
Afterwards, 53 electronics were built and are acquiring data since February 2018. After several tests, this design will be produced to complete the AMIGA muon counters, by installing three \SI{10}{\metre\squared} modules in 61 stations.

\acknowledgments

\begin{sloppypar}
The successful installation, commissioning, and operation of the Pierre Auger Observatory would not have been possible without the strong commitment and effort from the technical and administrative staff in Malarg\"ue. We are very grateful to the following agencies and organizations for financial support:
\end{sloppypar}

\begin{sloppypar}
Comisi\'on Nacional de Energ\'\i{}a At\'omica, Agencia Nacional de Promoci\'on Cient\'\i{}fica y Tecnol\'ogica (ANPCyT), Consejo Nacional de Investigaciones Cient\'\i{}ficas y T\'ecnicas (CONICET), Gobierno de la Provincia de Mendoza, Municipalidad de Malarg\"ue, NDM Holdings and Valle Las Le\~nas, in gratitude for their continuing cooperation over land access, Argentina; the Australian Research Council; Conselho Nacional de Desenvolvimento Cient\'\i{}fico e Tecnol\'ogico (CNPq), Financiadora de Estudos e Projetos (FINEP), Funda\c{c}\~ao de Amparo \`a Pesquisa do Estado de Rio de Janeiro (FAPERJ), S\~ao Paulo Research Foundation (FAPESP) Grants No.\ 2010/07359-6 and No.\ 1999/05404-3, Minist\'erio de Ci\^encia e Tecnologia (MCT), Brazil; Grant No.\ MSMT CR LG15014, LO1305 and LM2015038 and the Czech Science Foundation Grant No.\ 14-17501S, Czech Republic; Centre de Calcul IN2P3/CNRS, Centre National de la Recherche Scientifique (CNRS), Conseil R\'egional Ile-de-France, D\'epartement Physique Nucl\'eaire et Corpusculaire (PNC-IN2P3/CNRS), D\'epartement Sciences de l'Univers (SDU-INSU/CNRS), Institut Lagrange de Paris (ILP) Grant No.\ LABEX ANR-10-LABX-63, within the Investissements d'Avenir Programme Grant No.\ ANR-11-IDEX-0004-02, France; Bundesministerium f\"ur Bildung und Forschung (BMBF), Deutsche Forschungsgemeinschaft (DFG), Finanzministerium Baden-W\"urttemberg, Helmholtz Alliance for Astroparticle Physics (HAP), Helmholtz-Gemeinschaft Deutscher Forschungszentren (HGF), Ministerium f\"ur Wissenschaft und Forschung, Nordrhein Westfalen, Ministerium f\"ur Wissenschaft, Forschung und Kunst, Baden-W\"urttemberg, Germany; Istituto Nazionale di Fisica Nucleare (INFN),Istituto Nazionale di Astrofisica (INAF), Ministero dell'Istruzione, dell'Universit\'a e della Ricerca (MIUR), Gran Sasso Center for Astroparticle Physics (CFA), CETEMPS Center of Excellence, Ministero degli Affari Esteri (MAE), Italy; Consejo Nacional de Ciencia y Tecnolog\'\i{}a (CONACYT) No.\ 167733, Mexico; Universidad Nacional Aut\'onoma de M\'exico (UNAM), PAPIIT DGAPA-UNAM, Mexico; Ministerie van Onderwijs, Cultuur en Wetenschap, Nederlandse Organisatie voor Wetenschappelijk Onderzoek (NWO), Stichting voor Fundamenteel Onderzoek der Materie (FOM), Netherlands; National Centre for Research and Development, Grants No.\ ERA-NET-ASPERA/01/11 and No.\ ERA-NET-ASPERA/02/11, National Science Centre, Grants No.\ 2013/08/M/ST9/00322, No.\ 2013/08/M/ST9/00728 and No.\ HARMONIA 5 -- 2013/10/M/ST9/00062, Poland; Portuguese national funds and FEDER funds within Programa Operacional Factores de Competitividade through Funda\c{c}\~ao para a Ci\^encia e a Tecnologia (COMPETE), Portugal; Romanian Authority for Scientific Research ANCS, CNDI-UEFISCDI partnership projects Grants No.\ 20/2012 and No.194/2012 and PN 16 42 01 02; Slovenian Research Agency, Slovenia; Comunidad de Madrid, Fondo Europeo de Desarrollo Regional (FEDER) funds, Ministerio de Econom\'\i{}a y Competitividad, Xunta de Galicia, European Community 7th Framework Program, Grant No.\ FP7-PEOPLE-2012-IEF-328826, Spain; Science and Technology Facilities Council, United Kingdom; Department of Energy, Contracts No.\ DE-AC02-07CH11359, No.\ DE-FR02-04ER41300, No.\ DE-FG02-99ER41107 and No.\ DE-SC0011689, National Science Foundation, Grant No.\ 0450696, The Grainger Foundation, USA; NAFOSTED, Vietnam; Marie Curie-IRSES/EPLANET, European Particle Physics Latin American Network, European Union 7th Framework Program, Grant No.\ PIRSES-2009-GA-246806; and UNESCO.
\end{sloppypar}


\bibliographystyle{JHEP}
\bibliography{Papers,Datasheets}{}

\begin{center}
\rule{0.1\columnwidth}{0.5pt}\,\raisebox{-0.5pt}{\rule{0.05\columnwidth}{1.5pt}}~\raisebox{-0.375ex}{\scriptsize$\bullet$}~\raisebox{-0.5pt}{\rule{0.05\columnwidth}{1.5pt}}\,\rule{0.1\columnwidth}{0.5pt}
\end{center}

\section*{The Pierre Auger Collaboration}

A.~Aab$^{80}$,
P.~Abreu$^{72}$,
M.~Aglietta$^{52,50}$,
J.M.~Albury$^{12}$,
I.~Allekotte$^{1}$,
A.~Almela$^{8,11}$,
J.~Alvarez-Mu\~niz$^{79}$,
R.~Alves Batista$^{80}$,
G.A.~Anastasi$^{61,50}$,
L.~Anchordoqui$^{87}$
% ...
{\footnotesize

\begin{description}[labelsep=0.2em,align=right,labelwidth=0.7em,labelindent=0em,leftmargin=2em,noitemsep]
\item[$^{1}$] Centro At\'omico Bariloche and Instituto Balseiro (CNEA-UNCuyo-CONICET), San Carlos de Bariloche, Argentina
\item[$^{2}$] Centro de Investigaciones en L\'aseres y Aplicaciones, CITEDEF and CONICET, Villa Martelli, Argentina
\item[$^{3}$] Departamento de F\'\i{}sica and Departamento de Ciencias de la Atm\'osfera y los Oc\'eanos, FCEyN, Universidad de Buenos Aires and CONICET, Buenos Aires, Argentina
\item[$^{4}$] IFLP, Universidad Nacional de La Plata and CONICET, La Plata, Argentina
\item[$^{5}$] Instituto de Astronom\'\i{}a y F\'\i{}sica del Espacio (IAFE, CONICET-UBA), Buenos Aires, Argentina
\item[$^{6}$] Instituto de F\'\i{}sica de Rosario (IFIR) -- CONICET/U.N.R.\ and Facultad de Ciencias Bioqu\'\i{}micas y Farmac\'euticas U.N.R., Rosario, Argentina
\item[$^{7}$] Instituto de Tecnolog\'\i{}as en Detecci\'on y Astropart\'\i{}culas (CNEA, CONICET, UNSAM), and Universidad Tecnol\'ogica Nacional -- Facultad Regional Mendoza (CONICET/CNEA), Mendoza, Argentina
\item[$^{8}$] Instituto de Tecnolog\'\i{}as en Detecci\'on y Astropart\'\i{}culas (CNEA, CONICET, UNSAM), Buenos Aires, Argentina
\item[$^{9}$] Observatorio Pierre Auger, Malarg\"ue, Argentina
\item[$^{10}$] Observatorio Pierre Auger and Comisi\'on Nacional de Energ\'\i{}a At\'omica, Malarg\"ue, Argentina
\item[$^{11}$] Universidad Tecnol\'ogica Nacional -- Facultad Regional Buenos Aires, Buenos Aires, Argentina
\item[$^{12}$] University of Adelaide, Adelaide, S.A., Australia
\item[$^{13}$] Universit\'e Libre de Bruxelles (ULB), Brussels, Belgium
\item[$^{14}$] Vrije Universiteit Brussels, Brussels, Belgium
\item[$^{15}$] Centro Brasileiro de Pesquisas Fisicas, Rio de Janeiro, RJ, Brazil
\item[$^{16}$] Centro Federal de Educa\c{c}\~ao Tecnol\'ogica Celso Suckow da Fonseca, Nova Friburgo, Brazil
\item[$^{17}$] Universidade de S\~ao Paulo, Escola de Engenharia de Lorena, Lorena, SP, Brazil
\item[$^{18}$] Universidade de S\~ao Paulo, Instituto de F\'\i{}sica de S\~ao Carlos, S\~ao Carlos, SP, Brazil
\item[$^{19}$] Universidade de S\~ao Paulo, Instituto de F\'\i{}sica, S\~ao Paulo, SP, Brazil
\item[$^{20}$] Universidade Estadual de Campinas, IFGW, Campinas, SP, Brazil
\item[$^{21}$] Universidade Estadual de Feira de Santana, Feira de Santana, Brazil
\item[$^{22}$] Universidade Federal do ABC, Santo Andr\'e, SP, Brazil
\item[$^{23}$] Universidade Federal do Paran\'a, Setor Palotina, Palotina, Brazil
\item[$^{24}$] Universidade Federal do Rio de Janeiro, Instituto de F\'\i{}sica, Rio de Janeiro, RJ, Brazil
\item[$^{25}$] Universidade Federal do Rio de Janeiro (UFRJ), Observat\'orio do Valongo, Rio de Janeiro, RJ, Brazil
\item[$^{26}$] Universidade Federal Fluminense, EEIMVR, Volta Redonda, RJ, Brazil
\item[$^{27}$] Universidad de Medell\'\i{}n, Medell\'\i{}n, Colombia
\item[$^{28}$] Universidad Industrial de Santander, Bucaramanga, Colombia
\item[$^{29}$] Charles University, Faculty of Mathematics and Physics, Institute of Particle and Nuclear Physics, Prague, Czech Republic
\item[$^{30}$] Institute of Physics of the Czech Academy of Sciences, Prague, Czech Republic
\item[$^{31}$] Palacky University, RCPTM, Olomouc, Czech Republic
\item[$^{32}$] CNRS/IN2P3, IJCLab, Universit\'e Paris-Saclay, Orsay, France
\item[$^{33}$] Laboratoire de Physique Nucl\'eaire et de Hautes Energies (LPNHE), Sorbonne Universit\'e, Universit\'e de Paris, CNRS-IN2P3, Paris, France
\item[$^{34}$] Univ.\ Grenoble Alpes, CNRS, Grenoble Institute of Engineering Univ.\ Grenoble Alpes, LPSC-IN2P3, 38000 Grenoble, France
\item[$^{35}$] Universit\'e Paris-Saclay, CNRS/IN2P3, IJCLab, Orsay, France
\item[$^{36}$] Bergische Universit\"at Wuppertal, Department of Physics, Wuppertal, Germany
\item[$^{37}$] Karlsruhe Institute of Technology (KIT), Institute for Experimental Particle Physics, Karlsruhe, Germany
\item[$^{38}$] Karlsruhe Institute of Technology (KIT), Institut f\"ur Prozessdatenverarbeitung und Elektronik, Karlsruhe, Germany
\item[$^{39}$] Karlsruhe Institute of Technology (KIT), Institute for Astroparticle Physics, Karlsruhe, Germany
\item[$^{40}$] RWTH Aachen University, III.\ Physikalisches Institut A, Aachen, Germany
\item[$^{41}$] Universit\"at Hamburg, II.\ Institut f\"ur Theoretische Physik, Hamburg, Germany
\item[$^{42}$] Universit\"at Siegen, Department Physik -- Experimentelle Teilchenphysik, Siegen, Germany
\item[$^{43}$] Gran Sasso Science Institute, L'Aquila, Italy
\item[$^{44}$] INFN Laboratori Nazionali del Gran Sasso, Assergi (L'Aquila), Italy
\item[$^{45}$] INFN, Sezione di Catania, Catania, Italy
\item[$^{46}$] INFN, Sezione di Lecce, Lecce, Italy
\item[$^{47}$] INFN, Sezione di Milano, Milano, Italy
\item[$^{48}$] INFN, Sezione di Napoli, Napoli, Italy
\item[$^{49}$] INFN, Sezione di Roma ``Tor Vergata'', Roma, Italy
\item[$^{50}$] INFN, Sezione di Torino, Torino, Italy
\item[$^{51}$] Istituto di Astrofisica Spaziale e Fisica Cosmica di Palermo (INAF), Palermo, Italy
\item[$^{52}$] Osservatorio Astrofisico di Torino (INAF), Torino, Italy
\item[$^{53}$] Politecnico di Milano, Dipartimento di Scienze e Tecnologie Aerospaziali , Milano, Italy
\item[$^{54}$] Universit\`a del Salento, Dipartimento di Matematica e Fisica ``E.\ De Giorgi'', Lecce, Italy
\item[$^{55}$] Universit\`a dell'Aquila, Dipartimento di Scienze Fisiche e Chimiche, L'Aquila, Italy
\item[$^{56}$] Universit\`a di Catania, Dipartimento di Fisica e Astronomia, Catania, Italy
\item[$^{57}$] Universit\`a di Milano, Dipartimento di Fisica, Milano, Italy
\item[$^{58}$] Universit\`a di Napoli ``Federico II'', Dipartimento di Fisica ``Ettore Pancini'', Napoli, Italy
\item[$^{59}$] Universit\`a di Palermo, Dipartimento di Fisica e Chimica ''E.\ Segr\`e'', Palermo, Italy
\item[$^{60}$] Universit\`a di Roma ``Tor Vergata'', Dipartimento di Fisica, Roma, Italy
\item[$^{61}$] Universit\`a Torino, Dipartimento di Fisica, Torino, Italy
\item[$^{62}$] Benem\'erita Universidad Aut\'onoma de Puebla, Puebla, M\'exico
\item[$^{63}$] Centro de Investigaci\'on y de Estudios Avanzados del IPN (CINVESTAV), M\'exico, D.F., M\'exico
\item[$^{64}$] Unidad Profesional Interdisciplinaria en Ingenier\'\i{}a y Tecnolog\'\i{}as Avanzadas del Instituto Polit\'ecnico Nacional (UPIITA-IPN), M\'exico, D.F., M\'exico
\item[$^{65}$] Universidad Aut\'onoma de Chiapas, Tuxtla Guti\'errez, Chiapas, M\'exico
\item[$^{66}$] Universidad Michoacana de San Nicol\'as de Hidalgo, Morelia, Michoac\'an, M\'exico
\item[$^{67}$] Universidad Nacional Aut\'onoma de M\'exico, M\'exico, D.F., M\'exico
\item[$^{68}$] Universidad Nacional de San Agustin de Arequipa, Facultad de Ciencias Naturales y Formales, Arequipa, Peru
\item[$^{69}$] Institute of Nuclear Physics PAN, Krakow, Poland
\item[$^{70}$] University of \L{}\'od\'z, Faculty of Astrophysics, \L{}\'od\'z, Poland
\item[$^{71}$] University of \L{}\'od\'z, Faculty of High-Energy Astrophysics,\L{}\'od\'z, Poland
\item[$^{72}$] Laborat\'orio de Instrumenta\c{c}\~ao e F\'\i{}sica Experimental de Part\'\i{}culas -- LIP and Instituto Superior T\'ecnico -- IST, Universidade de Lisboa -- UL, Lisboa, Portugal
\item[$^{73}$] ``Horia Hulubei'' National Institute for Physics and Nuclear Engineering, Bucharest-Magurele, Romania
\item[$^{74}$] Institute of Space Science, Bucharest-Magurele, Romania
\item[$^{75}$] University Politehnica of Bucharest, Bucharest, Romania
\item[$^{76}$] Center for Astrophysics and Cosmology (CAC), University of Nova Gorica, Nova Gorica, Slovenia
\item[$^{77}$] Experimental Particle Physics Department, J.\ Stefan Institute, Ljubljana, Slovenia
\item[$^{78}$] Universidad de Granada and C.A.F.P.E., Granada, Spain
\item[$^{79}$] Instituto Galego de F\'\i{}sica de Altas Enerx\'\i{}as (IGFAE), Universidade de Santiago de Compostela, Santiago de Compostela, Spain
\item[$^{80}$] IMAPP, Radboud University Nijmegen, Nijmegen, The Netherlands
\item[$^{81}$] KVI -- Center for Advanced Radiation Technology, University of Groningen, Groningen, The Netherlands
\item[$^{82}$] Nationaal Instituut voor Kernfysica en Hoge Energie Fysica (NIKHEF), Science Park, Amsterdam, The Netherlands
\item[$^{83}$] Stichting Astronomisch Onderzoek in Nederland (ASTRON), Dwingeloo, The Netherlands
\item[$^{84}$] Universiteit van Amsterdam, Faculty of Science, Amsterdam, The Netherlands
\item[$^{85}$] Case Western Reserve University, Cleveland, OH, USA
\item[$^{86}$] Colorado School of Mines, Golden, CO, USA
\item[$^{87}$] Department of Physics and Astronomy, Lehman College, City University of New York, Bronx, NY, USA
\item[$^{88}$] Louisiana State University, Baton Rouge, LA, USA
\item[$^{89}$] Michigan Technological University, Houghton, MI, USA
\item[$^{90}$] New York University, New York, NY, USA
\item[$^{91}$] Pennsylvania State University, University Park, PA, USA
\item[$^{92}$] University of Chicago, Enrico Fermi Institute, Chicago, IL, USA
\item[$^{93}$] University of Delaware, Department of Physics and Astronomy, Bartol Research Institute, Newark, DE, USA
\item[$^{94}$] University of Wisconsin-Madison, Department of Physics and WIPAC, Madison, WI, USA
\item[] -----
\item[$^{a}$] Fermi National Accelerator Laboratory, Fermilab, Batavia, IL, USA
\item[$^{b}$] Max-Planck-Institut f\"ur Radioastronomie, Bonn, Germany
\item[$^{c}$] School of Physics and Astronomy, University of Leeds, Leeds, United Kingdom
\item[$^{d}$] Colorado State University, Fort Collins, CO, USA
\item[$^{e}$] now at Hakubi Center for Advanced Research and Graduate School of Science, Kyoto University, Kyoto, Japan
\item[$^{f}$] also at University of Bucharest, Physics Department, Bucharest, Romania
\end{description}

}
\end{document}